\documentclass[letter,11pt,final]{article}
\usepackage{typearea}
\typearea{14}
\makeatletter

\usepackage{fullpage}

\usepackage{authblk}
\usepackage{amsmath, amssymb, amsthm, thmtools, amsfonts, bm}
\usepackage{dsfont}
\usepackage{nicefrac}
\usepackage{cite}
\usepackage[numbers,sort]{natbib}
\usepackage{url}
\usepackage[usenames,dvipsnames]{xcolor}
\usepackage{algorithm}
\usepackage{algorithmicx}
\usepackage[noend]{algpseudocode}
\usepackage{epstopdf}
\usepackage[shortlabels]{enumitem}
\usepackage{mleftright}

\usepackage{framed}
\usepackage[framemethod=tikz]{mdframed}
\usepackage{xspace}

\usepackage[colorlinks,citecolor=blue,linkcolor=BrickRed]{hyperref}
\usepackage{cleveref}

\theoremstyle{plain}
\newtheorem{thm}{Theorem}[section]
\newtheorem{theorem}{Theorem}[section]

\newtheorem{corollary}[thm]{Corollary}
\newtheorem{property}[thm]{Property}
\newtheorem{invariant}[thm]{Invariant}

\newtheorem{lemma}[thm]{Lemma}
\newtheorem{definition}[thm]{Definition}
\newtheorem{observation}[thm]{Observation}
\newtheorem{claim}[thm]{Claim}

\title{Deterministic Rounding of Dynamic Fractional Matchings}
\author[1]{Sayan Bhattacharya\footnote{Supported by Engineering and Physical Sciences Research Council, UK (EPSRC)  Grant  EP/S03353X/1.}}
\author[2]{Peter Kiss}
\affil[1]{University of Warwick}
\affil[2]{University of Warwick}
\date{}

\begin{document}

\begin{titlepage}
\maketitle
\pagenumbering{gobble}

\begin{abstract}

We present a framework for {\em deterministically}  rounding a dynamic fractional matching. Applying our framework in a black-box manner on top of existing fractional matching algorithms, we derive the following new results:  (1) The first deterministic algorithm for maintaining a $(2-\delta)$-approximate maximum matching in a fully dynamic bipartite graph, in arbitrarily small polynomial update time. (2) The first deterministic algorithm for maintaining a $(1+\delta)$-approximate maximum matching in a decremental bipartite graph, in polylogarithmic  update time. (3) The first deterministic algorithm for maintaining a $(2+\delta)$-approximate maximum matching in a fully dynamic general  graph, in {\em small} polylogarithmic (specifically, $O(\log^4 n)$) update time. These results are respectively obtained by applying our framework on top of the fractional matching algorithms of Bhattacharya et al.~[STOC'16], Bernstein et al.~[FOCS'20], and Bhattacharya and Kulkarni~[SODA'19]. 

Prior to our work, there were two known general-purpose rounding schemes for dynamic fractional matchings. Both these  schemes, by Arar et al.~[ICALP'18] and Wajc~[STOC'20], were randomized. 

Our  rounding scheme works by maintaining a good {\em matching-sparsifier} with bounded arboricity, and then applying the algorithm of Peleg and Solomon [SODA'16] to maintain a near-optimal matching in this low arboricity graph. To the best of our knowledge, this is the first dynamic matching algorithm that works on general graphs by using an algorithm for low-arboricity graphs as a black-box subroutine. This feature of our rounding scheme might be of independent interest.

\end{abstract}

\newpage

	\setcounter{tocdepth}{2}
	\tableofcontents

\end{titlepage}
\pagenumbering{arabic}

\section{Introduction}
The central question in the area of {\em dynamic algorithms} is to understand how can we efficiently maintain a good solution to a computational problem, when the underlying input changes over time~\cite{HenzingerK95,HolmLT01}. In the past decade, an extensive body of work in this area has been devoted to the study of {\em dynamic matching}~\cite{AbboudA0PS19,BaswanaGS11, BehnezhadDHSS19, BehnezhadLM20, BernsteinFH19, BernsteinS15, BernsteinS16, CharikarS18, GrandoniLSSS19, Gupta14, NeimanS13, OnakR10, Sankowski07, Solomon16}. 

A matching $M \subseteq E$ in $G$ is a set of edges that do not share any common endpoint. In the dynamic matching problem,  the input is a  graph $G = (V, E)$ that keeps getting updated via edge insertions/deletions, and the goal is to maintain an approximately {\em maximum matching} in $G$ with small (preferably polylogarithmic) update time, where the phrase ``update time'' refers to the time it takes to handle an ``update'' (edge insertion/deletion) in $G$.\footnote{An algorithm has an ``amortized'' update time of $O(\tau)$ iff starting from an empty graph, it can handle any sequence of $\kappa$ edges insertions/deletions in $O(\tau \cdot \kappa)$ total time.} From the current landscape of dynamic matching, we can  identify a common template that underpins a number of existing  algorithms for this problem. This template consists of three  steps. 

\smallskip
\noindent 
{\bf Step (I):} Design an efficient dynamic algorithm that maintains an approximately maximum {\em fractional matching}\footnote{A fractional matching $w$ in $G$ assigns a weight $w(e) \in [0, 1]$ to every edge $e \in E$, ensuring that the total weight assigned to all the edges incident on any given node is $\leq 1$.} $w : E \rightarrow [0, 1]$ in the input graph $G = (V, E)$. All the known algorithms for this first step are deterministic~\cite{BernsteinGS20, BhattacharyaCH17, BhattacharyaHI15, BhattacharyaHN17, BhattacharyaHN16, BhattacharyaK19, GuptaKPP17}. 

\smallskip
\noindent
{\bf Step (II):} Maintain a {\em sparse} (bounded-degree) subgraph $S = (V, E_S)$ of the input graph, with $E_S \subseteq E$, that approximately preserves the size of maximum matching~\cite{ArarCCSW18,Wajc20}. In a bit more details, the subgraph $S$ should have the property that $\mu(S)$ is very close to $size(w)$, where $\mu(S)$ denotes the size of maximum (integral) matching in $S$, and $size(w) = \sum_{e \in E} w(e)$ denotes the size of the fractional matching $w$ from the previous step. Such a subgraph $S$ is often referred to as a {\em matching-sparsifier} of $G$~\cite{AssadiB19}. There are two known algorithms for this second step and both of them are randomized, in sharp contrast to  Step (I). Specifically, Arar et al.~\cite{ArarCCSW18} designed a randomized rounding scheme for sparsifying a dynamic fractional matching. Their algorithm works only in the {\em oblivious adversary} setting, where the future updates cannot depend on the past actions taken by the algorithm. This result was very recently improved upon by Wajc~\cite{Wajc20}, who presented an elegant dynamic rounding scheme for Step (II) that, although randomized, works in a much more general {\em adaptive adversary} setting, where the future updates to the algorithm can depend on all its past random bits. 

\smallskip
\noindent {\bf Step (III):} Maintain a near-optimal matching in the  (bounded-degree) sparsifier $S$ from the previous step, using a known algorithm by Gupta et al.~\cite{GuptaP13}, which has $O(\Delta)$  update time on dynamic graphs with maximum degree $\leq \Delta$. Since $S$ has bounded degree, the  third step incurs only a small overhead in the update time. The algorithm in~\cite{GuptaP13} is also deterministic. 

\smallskip
  A natural question arises from the preceding discussion. Can we design an efficient {\em deterministic} dynamic algorithm for Step (II)? Since Step (I) and Step (III) are already deterministic, an efficient deterministic algorithm for Step (II) will allow us to derandomize multiple existing results in the literature on dynamic matching. We resolve this question in the affirmative. Specifically, our  main result is summarized in the theorem below.

\begin{theorem}
\label{th:inf:main}
Fix any small constant $\delta > 0$. Consider a dynamic graph $G = (V, E)$ on $n$ nodes and a (dynamic) fractional matching  $w$ in $G$. In this setting, an update either inserts/deletes an edge in $G = (V, E)$ or changes the weight $w(e)$ of an existing edge $e \in E$. We can deterministically maintain a subgraph $S = (V, E_S)$ of $G$, with $E_S \subseteq E$, such that: 
\begin{enumerate}
\item There exists a fractional matching $h' : E_S \rightarrow [0, 1]$ in $S$ with $size(w) \leq (1+\delta) \cdot size(h')$.
\item If $w$ is a $(\delta, \delta)$-approximate maximal  matching in $G$, then $\mu(G) \leq (2+\delta) \cdot \mu(S)$.
\item  The arboricity of $S$ is $O(\log^2 n)$. 
\item Every update in $G$ or $w$, on average, leads to $O(\log^2 n)$ updates in $S$.
\item Our dynamic algorithm for maintaining $S$ has $O(\log^2 n)$ amortized update time.
\end{enumerate}
\end{theorem}

\noindent
{\bf Bounded arboricity matching-sparsifiers:} We will shortly explain  part-(2) of \Cref{th:inf:main}, which uses the notion of a {\em $(\delta, \delta)$-approximate maximal  matching} that has not been defined yet. For now, we focus on an intriguing feature of \Cref{th:inf:main}, namely,  that it only maintains a subgraph $S$ with bounded {\em arboricity}.\footnote{Informally, an undirected graph $G' = (V', E')$ has arboricity $\kappa$ if we can assign a direction to each of its edges $e \in E'$  in such a way that every node $v \in V'$ gets an out-degree of at most $O(\kappa)$. If a graph has maximum degree at most $\kappa$, then its arboricity is also $O(\kappa)$, but not vice versa.} This is in sharp contrast to all previous work on dynamic matching-sparsifiers: they satisfy the strictly stronger requirement of bounded maximum degree~\cite{ArarCCSW18,Wajc20}. Our algorithm exploits this feature in a crucial manner, allowing certain nodes to have large degrees in $S$ while ensuring that the arboricity of $S$ remains at most $O(\log^2 n)$. This does not cause any problem in the overall scheme of things, however, because Peleg and Solomon~\cite{PelegS16} have shown how to deterministically maintain a $(1+\delta)$-approximate maximum matching in $O(\Delta)$ update time in a dynamic graph with arboricity $\leq \Delta$. Their algorithm allows us to efficiently maintain a near-optimal matching in $S$. 

To summarize, there is an existing line of work on  dynamic matching which deal with the special class of  low-arboricity graphs~\cite{KopelowitzKPS14,NeimanS13,PelegS16}. \Cref{th:inf:main} shows that if we have a good dynamic matching algorithm for low-arboricity graphs, then we can use it in a black-box manner to design better dynamic matching algorithms for {\em general graphs} as well.

\smallskip
\noindent {\bf Implications of \Cref{th:inf:main}:} We start by focussing  on  bipartite graphs. If a graph $G$ is bipartite, then the size of a maximum fractional matching in $G$ is equal to $\mu(G)$. Accordingly, part-(1) of \Cref{th:inf:main} implies that if the input graph $G$ is bipartite, then our dynamic algorithm  maintains a sparsifier $S = (V, E_S)$ such that $size(w) \leq (1+\delta) \cdot \mu(S)$. We can now run the dynamic algorithm from~\cite{PelegS16} on $S$, which has small arboricity, to efficiently maintain a near-optimal (integral) matching $M \subseteq E_S$ with $size(w) \leq (1+\delta) \cdot |M|$. 

Bhattacharya et al.~\cite{BhattacharyaHN16} gave a deterministic algorithm for maintaining  $(2-\epsilon)$-approximate maximum {\em fractional} matchings in  bipartite graphs with  arbitrarily small polynomial update time. Applying our  dynamic rounding framework on top of this result from~\cite{BhattacharyaHN16}, we get the {\em first deterministic} algorithm for dynamic (integral) matchings in bipartite graphs with the same approximation ratio and similar update time, as summarized in the theorem below.

\begin{theorem}
\label{th:apply:bipartite:1}
For every constant $k \geq 10$, there exists a $\beta_k \in (1, 2)$, and a {\em deterministic} dynamic algorithm that maintains a $\beta_k$-approximate maximum matching in an $n$-node bipartite graph with $O(n^{1/k} \cdot \log^4 n)$ amortized update time. 
\end{theorem}

Next, very recently Bernstein et al.~\cite{BernsteinGS20} showed how to maintain a $(1+\delta)$-approximate maximum {\em fractional} matching in a bipartite graph with $O(\log^3 n)$ amortized update time in the {\em decremental} setting, where the input graph only undergoes edge-deletions.  Applying our dynamic rounding framework on top of their result, we get the {\em first deterministic} algorithm for maximum (integral) matching in an analogous decremental setting, with the same approximation ratio and similar update time. This is stated in the theorem below.

\begin{theorem}
\label{th:apply:bipartite:2}
We can {\em deterministically} maintain  a $(1+\delta)$-approximate maximum matching in a decremental bipartite graph on $n$ nodes with $O(\log^7 n)$ amortized update time.
\end{theorem}

Moving on to general graphs, we note that if  a graph $G = (V, E)$ is non-bipartite, then the size of a maximum fractional matching can be as large as $(3/2) \cdot \mu(G)$. Thus, if we are to naively apply our dynamic rounding framework based on the guarantee given to us by part-(1) of \Cref{th:inf:main}, then we will lose out on a factor of $3/2$ in the approximation ratio.  This is where part-(2) of \Cref{th:inf:main} comes in handy. Specifically, as in~\cite{ArarCCSW18,Wajc20}, we invoke  the notion of an {\em $(\alpha, \beta)$-approximate maximal matching} (see \Cref{def:beta:approx}). 

To see why this notion is useful for us, consider the result of Bhattacharya and Kulkarni~\cite{BhattacharyaK19}, who designed a deterministic dynamic algorithm for $(2+\delta)$-approximate maximum {\em fractional} matching, for small constant $\delta > 0$, in general graphs with $O(1)$ amortized update time. Furthermore, the fractional matching maintained by~\cite{BhattacharyaK19} is $(\delta, \delta)$-approximately maximal. Thus, applying \Cref{th:inf:main} on top of this result from~\cite{BhattacharyaK19}, we can deterministically maintain a sparsifier $S = (V, E_S)$ of the input graph $G$ with $\mu(G) \leq (2+\delta) \cdot \mu(S)$. We can now maintain a near-optimal maximum matching $M \subseteq E_S$ in $S$, using the algorithm of~\cite{PelegS16}. Since $\mu(G) \leq (2+\delta) \cdot \mu(S)$, $M$ will be a $(2+\delta)$-approximate maximum (integral) matching in $G$. Putting everything together, we get the result summarized in the theorem below.

\begin{theorem}
\label{th:apply:general}
We can {\em deterministically} maintain a $(2+\delta)$-approximate maximum matching in an $n$-node dynamic graph with $O(\log^4 n)$ amortized update time, for small constant $\delta > 0$.
\end{theorem}

Prior to our work, the only {\em deterministic} dynamic algorithm for $(2+\delta)$-approximate maximum matching in general graphs with polylogarithmic update time was due to Bhattacharya et al.~\cite{BhattacharyaHN16}. The exact polylogarithmic factor in the update time of~\cite{BhattacharyaHN16} was huge (more than $\log^{20} n$), and the algorithm of~\cite{BhattacharyaHN16} was significantly more complicated than ours.

\smallskip
\noindent {\bf Perspective:}  Existing techniques for proving  update-time lower bounds for dynamic problems cannot distinguish between deterministic and randomized algorithms~\cite{AbboudW14, HenzingerKNS15, KopelowitzPP16, Larsen12, Patrascu10}. Thus, understanding the power of randomization in the dynamic setting is an important research agenda,  which comprises  of two separate strands of work. (1) Studying the power of the {\em oblivious adversary} assumption while designing a randomized algorithm for a given dynamic problem.  (2) Studying the separation between  randomized algorithms that work against {\em adaptive adversaries} on the one hand, and deterministic algorithms on the other. Our work falls under the second category.  A recent breakthrough result under this  category has been a deterministic algorithm for dynamic minimum spanning forest with worst-case subpolynomial update time~\cite{ChuzhoyGLNPS20,GoranciRST21}. This improves upon earlier work which achieved the same update time guarantee for dynamic minimum spanning forest, but using a randomized algorithm that works against adaptive adversary~\cite{NanongkaiSW17}. There are other well-studied dynamic problems where currently we have  polynomial gaps between the update times of the best-known deterministic algorithm and the best-known randomized algorithm against adaptive adversary~\cite{DuanHZ19, EvaldHGN20}. Bridging these gaps remain challenging open questions.

\smallskip
\noindent {\bf Our Techniques:} The key ingredient in our rounding scheme is a simple {\em degree-split} procedure. Given any graph $G' = (V', E')$ as input, this procedure runs in linear time and outputs a subgraph $G'' = (V', E'')$ where the degree of every node $v \in V'$ drops by a factor of $(1/2) \cdot (1\pm \epsilon)$, provided the initial degree of $v$ in $G'$ was larger than $(1/\epsilon)$. See \Cref{alg:degree-split}.

Using this degree-split procedure, we first design a simple {\em static} algorithm for sparsifying a {\em uniform} fractional  matching $w$ (which assigns the same weight to every edge) in an input graph $G = (V, E)$. This  works in rounds. In each round, we start by repeatedly removing the  nodes with degree at most $(1/\epsilon)$, until we are left with a graph $G' = (V', E')$ where every remaining node has degree larger than $(1/\epsilon)$. We now apply the degree-split procedure on $G' = (V', E')$ to obtain a subset of edges $E'' \subseteq E'$, double the weight of every edge $e \in E''$, and discard the edges $e \in E'' \setminus E'$ from the support of the fractional matching. Since the degree-split procedure reduces the degree of every node in $V'$ by (approximately) a factor of $1/2$, it follows that we (approximately) preserve the total weight received by every node while implementing a given round. We can show that if we continue with this process for (roughly) logarithmic many rounds, then we end up with a bounded-arboricity subgraph of  the input graph $G$ that approximately preserves the size of the fractional matching $w$. In the dynamic setting, we try to mimic this static algorithm in a natural lazy manner. 

When the input is a dynamic graph $G$ and a (not necessarily uniform) fractional matching $w$, then, roughly speaking, we first discretize $w$ and then decompose it into $O(\log n)$ many uniform fractional matchings,  defined on mutually edge-disjoint subgraphs of $G$. We  run a dynamic algorithm for sparsifying a uniform fractional matching on each of these subgraphs, and we maintain the union of the outputs of all these $O(\log n)$ many dynamic sparsifiers.

\section{Notations and Preliminaries}

Throughout this paper, we  let $G = (V, E)$ denote the input graph, and $n = |V|$ will be the number of nodes in $G$. For any $v \in V$, $E' \subseteq E$ and $V' \subseteq V$, we  let $E'(v, V') = \{ (u, v) \in E' : u \in V'\}$ denote the set of edges in $E'$ that are incident on $v$ and have their other endpoints in $V'$.  To ease notations, we  define $E'(v) := E'(v, V)$. We  also define $\text{deg}_{E'}(v, V') := |E'(v, V')|$ and $\text{deg}_{E'}(v) := |E'(v)|$. Furthermore, given any subset of edges $E' \subseteq E$, we let $V(E') = \bigcup_{(u, v) \in E'} \{u, v\}$ denote the set of endpoints of the edges in $E'$. Throughout the rest of this paper, we will  consider $\delta$ to be some small constant, and we will fix two more parameters $\beta$ and $\epsilon$ as stated below.
\begin{eqnarray}
	10^{-3}  \geq  \delta  =  20  \cdot \beta  =  5000 \cdot \epsilon \cdot \log n  >  0. \label{eq:bound:delta} \label{eq:bound:epsilon:new} \label{eq:bound:beta}
\end{eqnarray}
Given any subset of edges $E' \subseteq E$, a {\em weight-function} $w' : E' \rightarrow [0, 1]$ assigns a (possibly fractional) weight $0 \leq w'(e) \leq 1$ to every edge $e \in E'$. We  say that $E'$ is the {\em support} of $w'$ and write $\text{{\sc Support}}(w') := E'$. The {\em size} of the weight-function $w'$ is defined as $size(w') := \sum_{e \in E'} w'(e)$. For any node $v \in V$, let $w'(v) := \sum_{(u, v) \in E'} w'(u, v)$ denote the total weight received by $v$ from all its incident edges under the weight-function $w'$. We say that $w'$ is a {\em fractional matching} in the graph $G' := (V, E')$ iff $w'(v) \leq 1$ for all $v \in V$. Since $E' \subseteq E$, we often abuse notation to say that such a weight-function $w'$ is a fractional matching in $G = (V, E)$ as well. For any $0 \leq \lambda \leq 1$, we say that $w'$ is a {\em $\lambda$-uniform} weight-function (or, fractional matching, if $w'(v) \leq 1$ for all $v \in V$) iff $w'(e) = \lambda$ for all edges $e \in E'$.

Let $\mu(G')$ and $\mu_{f}(G')$ respectively denote the size of  maximum matching and the size of  maximum fractional matching in a graph $G'$.  We will use the following well-known theorem. 

\begin{theorem}
\label{th:int:gap}
Consider any graph $G'$. If $G'$ is bipartite, then $\mu(G') = \mu_f(G')$. Otherwise, we have $\mu(G) \leq \mu_f(G') \leq (3/2) \cdot \mu(G)$.
\end{theorem}

We will use the notion of an {\em approximately maximal} matching as in Arar et al.~\cite{ArarCCSW18}.

\begin{definition}
\label{def:beta:approx}
Consider any graph $G' = (V', E')$ and a fractional matching $w'$ in $G'$. We say that $w'$ is a $(\alpha,\beta)$-approximately maximal matching in $G'$ iff the following  holds. For every edge $(u, v) \in E'$, either (1) \{$w'(u, v) \geq \beta$\}, or (2) \{there is at least one endpoint $x \in \{u, v\}$ such that $w'(x) \geq 1-\alpha$ and $w'(x, y) < \beta$ for all edges $(x, y) \in E'$ incident on $x$\}.
\end{definition}

An {\em orientation} of a graph $G' = (V', E')$ assigns a direction to every edge $(u, v) \in E'$. For the rest of this paper, whenever we say that a graph $G'$ has {\em arboricity} $O(\kappa)$, we mean that $G'$ admits an orientation of its edges where the maximum out-degree of a node is $O(\kappa)$~\cite{Nash61}.

\section{Static Algorithm for Sparsifying a Uniform Fractional Matching}
\label{sec:static}

\newcommand{\A}{\mathcal{A}}
\newcommand{\R}{\mathcal{R}}

In this section, we present a simple static algorithm for sparsifying a uniform fractional matching. This will form the basis of our dynamic algorithm in \Cref{sec:dyn:uniform} and \Cref{sec:dyn:general}.

As input, we receive a graph $G = (V, E)$ and a $\lambda$-uniform fractional matching $w : E \rightarrow [0, 1]$ in $G$, for some $\lambda \in [\delta / n^2, \beta)$. Define $L = L(\lambda)$ to be the largest integer $k$ such that $\beta/2 \leq 2^k \lambda < \beta$. Since $\delta$ is a constant and $\delta/n^2 \leq \lambda < \beta$, from~(\ref{eq:bound:beta}) we infer that $L = O(\log n)$.

The algorithm proceeds in {\em rounds} $i \in \{0, \ldots, L-1\}$. Before the start of round $i = 0$, we initialize $E^{(\geq 0)} := E$, $V^{(\geq 0)} := V$, $G^{(\geq 0)} := \left(V^{(\geq 0)}, E^{(\geq 0)}\right)$ and $\gamma^{(0)} := w$. Thus,  $\gamma^{(0)}$ is a $\lambda$-uniform fractional matching in $G$. In each round $i$, we  identify a subset of edges $F^{(i)} \subseteq E^{(\geq i)}$ that  get {\em frozen}, in the sense that they are {\em not} considered in subsequent rounds. Define $\mathcal{F}^{(i)} := E^{(\geq i)} \bigcup_{j=0}^{i-1} F^{(j)}$ and $\mathcal{H}^{(i)} := (V, \mathcal{F}^{(i)})$ for all $i \in [0,L]$. The following invariant will be satisfied in the beginning of each round $i \in [0, L]$. The weight-function $\gamma^{(i)} : \mathcal{F}^{(i)} \rightarrow [0,1]$ ensures that $\gamma^{(i)}(v) \simeq w(v)$ for all  $v \in V$. Clearly, this invariant holds for $i = 0$.

\smallskip
\noindent {\bf Implementing a given round $i \in [0, L-1]$:} Initialize $G' = (V', E') := G^{(\geq i)}$.  There are two distinct {\em steps} in this round. During the first step, we  keep iteratively removing the nodes  with degree $\leq (1/\epsilon)$ from $G'$. Let $V^{(i)} \subseteq V^{(\geq i)}$ be the collection of nodes that get removed from $G'$ in this manner, and let $F^{(i)} \subseteq E^{(\geq i)}$ denote the set of edges incident on $V^{(i)}$. Define $V^{(\geq i+1)} := V^{(\geq i)} \setminus V^{(i)}$. At the end of this first step, the status of $G' = (V', E')$ is as follows: $V' = V^{(\geq i+1)}$ and $E' = E^{(\geq i)} \setminus F^{(i)}$. Intuitively, we can afford to remove the nodes $V^{(\geq i)}$ from $G'$ because the edges in $F^{(i)}$ admit an orientation with maximum out-degree $(1/\epsilon)$.  
At the end of this first step every node in $G'$ has degree $\geq (1/\epsilon)$. 

In the second step, we call a subroutine {\sc Degree-Split}$(E')$, which returns  a subset of edges $E'' \subseteq E'$ with the following property: $\text{deg}_{E''}(v) \simeq (1/2) \cdot \text{deg}_{E'}(v)$ for all nodes $v \in V'$. We will shortly see how to implement this {\sc Degree-Split} subroutine. For now, we move ahead with the description of round $i$. We set $E^{(\geq i+1)} := E''$ and $G^{(\geq i+1)} := \left(V^{(\geq i+1)}, E^{(\geq i+1)} \right)$. Next, we discard the edges in $E' \setminus E^{(\geq i+1)}$ from the support of $\gamma$ and double the weights on the remaining  edges in $E^{(\geq i+1)}$. This leads us to a new weight-function $\gamma^{(i+1)} : \mathcal{F}^{(i+1)} \rightarrow [0, 1]$ in $\mathcal{H}^{(i+1)} = (V, \mathcal{F}^{(i+1)})$ which is defined as follows. 
For every edge $e \in \mathcal{F}^{(\geq i+1)}$, we have:
\begin{equation}
\label{eq:text:1}
\gamma^{(i+1)}(e) =\begin{cases} 2 \cdot \gamma^{(i)}(e) & \text{ if } e \in E^{(\geq i+1)}; \\
\gamma^{(i)}(e) & \text{ else if } e \in \bigcup_{j = 0}^{i} F^{(j)}. 
\end{cases}
\end{equation}
At this point, if $i < L-1$, then we are ready to proceed to the next round $i+1$. Otherwise, if $i = L-1$, then  we terminate the algorithm after setting $F^{(L)} := E^{(\geq L)}$ and $V^{(L)} := V^{(\geq L)}$. 

\smallskip
Define $F := \bigcup_{i=0}^L F^{(i)}$,  $H := (V, F)$, and $h := \gamma^{(L)}$. We will show that $H = (V, F)$ is a good matching-sparsifier for the $\lambda$-uniform matching $w$ in the input graph $G = (V, E)$. 
The relevant pseudocodes  are summarized in  \Cref{alg:uniform-sparsify},  \Cref{alg:build}  and  \Cref{alg:degree-split}. For clarity of exposition,  in some of these pseudocodes we use one additional notation $h^{(i)}$, which is basically a weight-function $h^{(i)} : F^{(i)} \rightarrow [0, 1]$ such that $h^{(i)}(e) = 2^i \lambda$ for all $e \in F^{(i)}$.

\smallskip
\noindent {\bf Implementing the {\sc Degree-Split}$(E')$ subroutine:} Consider any graph $G^* = (V^*, E^*)$. A {\em walk} $W$ in $G^*$ is a set of distinct edges $\{ (u_0, v_0), \ldots, (u_{k}, v_{k})\} \subseteq E^*$ such that  $v_i = u_{i+1}$ for all $i \in [0, k-1]$. Let $W^{(even)} = \{ (u_{2i}, v_{2i}) : i \in [0, \lfloor k/2 \rfloor] \}$ denote the collection of even numbered edges from this walk $W$. The walk $W$ is said to be {\em maximal} in $G^*$ iff $\text{deg}_{E^* \setminus W}(u_0) = \text{deg}_{E^* \setminus W}(u_k) = 0$. When we call  {\sc Degree-Split}$(E')$, it first partitions the edge-set $E'$ into a collection of walks $\mathcal{W}$ as specified in  \Cref{alg:degree-split}. It then returns the set $E'' \subseteq E'$, which consists of all the even numbered edges from  all the walks $W \in \mathcal{W}$.

\begin{claim}
\label{cl:static:deg:split:1}
The subroutine {\sc Degree-Split}$(E')$ runs in $O(|E'|)$ time.
\end{claim}

\begin{proof}
We can compute a maximal walk $W$ in a given  graph $G^* = (V^*, E^*)$ in $O(|W|)$ time. Hence, the total running time of  \Cref{alg:degree-split} is given by $\sum_{W \in \mathcal{W}} O(\left| W \right|) = O(|E'|)$. 
\end{proof}

\begin{claim}
\label{cl:static:deg:split:2}
In  \Cref{alg:degree-split}, $\text{deg}_{E''}(v) \in \left[
\frac{\text{deg}_{E'}(v)}{2} - 1, \frac{\text{deg}_{E'}(v)}{2} + 1 \right]$ for all nodes $v \in V(E')$.
\end{claim}

\begin{proof}
Consider any graph $G^* = (V^*, E^*)$ and a walk $W = \{ (u_0, v_0), \ldots, (u_k, v_k)\}$ in $G^*$. Let $\text{{\sc End}}(W) = \{u_0, v_k\}$ denote the endpoints of this walk $W$, and let $V(W) = \bigcup_{i=0}^k \{u_i\} \bigcup_{i=0}^k \{v_i\}$ denote the set of all nodes touched by  $W$. Note that $\text{deg}_{W^{(even)}}(v) = (1/2) \cdot \text{deg}_W(v)$ for all nodes  $v \in V(W) \setminus \text{{\sc End}}(W)$, and $\text{deg}_{W^{(even)}}(v) \in \left[(1/2) \cdot \text{deg}_W(v) - 1, (1/2) \cdot \text{deg}_W(v) + 1\right]$ for all nodes $v \in \text{{\sc End}}(W)$. 
Now, fix any node $v \in V(E')$, and  observe that:
\begin{enumerate}
\item $\text{deg}_{E''}(v) = \sum_{W \in \mathcal{W}} \text{deg}_{W^{(even)}}(v)$ and $\text{deg}_{E'}(v) = \sum_{W \in \mathcal{W}} \text{deg}_{W}(v)$.
\item   The definition of a maximal walk implies that $v \in \text{{\sc End}}(W)$ for at most one  $W \in \mathcal{W}$.
\end{enumerate}
These observations, taken together, imply the claim.
\end{proof}

 \begin{algorithm}[H]
		\caption{{\sc Static-Uniform-Sparsify}$(G = (V, E),  \lambda)$, where $\delta / n^2 \leq \lambda < \beta$}.
		\label{alg:uniform-sparsify}
		\begin{algorithmic}
		        \State Let $w : E \rightarrow [0, 1]$ be a $\lambda$-uniform fractional matching in $G$. 
			\State Initialize $V^{(\geq 0)} := V$ and $E^{(\geq 0)} := E$.
			\State Initialize a weight-function $h^{(0)} : E^{(\geq 0)} \rightarrow [0, 1]$ so that $h^{(0)}(e) := \lambda$ for all $e \in E^{(\geq 0)}$.
			\State Let $L := L(\lambda)$ be the unique nonnegative integer $k$ such that $\beta/2 \leq 2^k \lambda < \beta$. 
			\State Call the subroutine {\sc Rebuild}$(0, \lambda)$. (see \Cref{alg:build})		
			\State Define $F := \bigcup_{i=0}^{L} F^{(i)}$, and $H := (V, F)$. 
			\State Define  $h : F \rightarrow [0, 1]$ such that for all $i \in [0, L]$ and $e \in F^{(i)}$ we have $h(e) := h^{(i)}(e)$.
		\end{algorithmic}
	\end{algorithm}

	\begin{algorithm}[H]
		\caption{{\sc Rebuild}$(i',    \lambda)$}
		\label{alg:build}
		\begin{algorithmic}
		       \For{$i=i'$ to $\left(L-1\right)$}:
			        \State $V^{(i)} \leftarrow \emptyset$.
				\While{there is some node $v \in V^{(\geq i)}\setminus V^{(i)}$ with $\text{deg}_{E^{(\geq i)}}\left(v, V^{(\geq i)} \setminus V^{(i)} \right) \leq (1/\epsilon)$}:
					\State $V^{(i)} \leftarrow V^{(i)} \cup \{v\}$.
				\EndWhile
				\State $V^{(\geq i+1)} \leftarrow V^{(\geq i)} \setminus V^{(i)}$. 
				\State $F^{(i)} \leftarrow \{ (u, v) \in E^{(\geq i)} :  \text{ either } u \in V^{(i)} \text{ or } v \in V^{(i)} \}$. 
				\State $E^{(\geq i+1)} \leftarrow \text{{\sc Degree-Split}}(E^{(\geq i)} \setminus F^{(i)})$.
				\For{all edges $e \in E^{(\geq i+1)}$}:
					\State $h^{(i+1)}(e) \leftarrow 2 \cdot h^{(i)}(e)$. 
				\EndFor
			\EndFor
			\State $F^{(L)} \leftarrow E^{(\geq L)}$. 		
			\State $V^{(L)} \leftarrow V^{(\geq L)}$. 
		\end{algorithmic}
	\end{algorithm}

	\begin{algorithm}[H]
	\caption{{\sc Degree-Split}$(E')$}
	\label{alg:degree-split}
		\begin{algorithmic}
			\State Initialize $E^* \leftarrow E'$ and $\mathcal{W} \leftarrow \emptyset$.
			\While{$E^* \neq \emptyset$}
			\State Let $G^* := (V(E^*), E^*)$, where $V(E^*)$ is the set of endpoints of the edges in $E^*$.
			\State Compute a {\em maximal} walk $W$ in $G^*$.
			\State Set $\mathcal{W} \leftarrow \mathcal{W} \cup \{ W \}$, and $E^* \leftarrow E^* \setminus W$.
			\EndWhile
			\State Return the set of edges $E'' := \bigcup_{W \in \mathcal{W}} W^{(even)}$. 
		\end{algorithmic}
	\end{algorithm}

Note that $h^{(i)}$ and $F^{(i)}$  represent global variables in the pseudocodes above.

\begin{lemma}
\label{lm:static:runtime}
 \Cref{alg:uniform-sparsify} runs in $O(|E|)$ time.
\end{lemma}

\begin{proof}
The runtime of  \Cref{alg:uniform-sparsify} is dominated by the call to {\sc Rebuild}$(0, \lambda)$. Accordingly, focus on any given iteration $i \in [0, L-1]$ of the outer {\sc For} loop in  \Cref{alg:build}.  During this iteration, using appropriate data structures the inner {\sc While} loop takes $O(|F^{(i)}|)$ time, the call to {\sc Degree-Split}$(E^{(\geq i)} \setminus F^{(i)})$ takes $O(|E^{(\geq i)} \setminus F^{(i)}|)$ time as per  \Cref{cl:static:deg:split:1}, and the inner {\sc For} loop takes $O(|E^{(\geq i+1)}|)$ time. Hence, the total time taken to implement iteration $i$ of the outer {\sc For} loop is at most $O(|E^{(\geq i)}|) + O(|E^{(\geq i)} \setminus F^{(i)}|) + O(|E^{(\geq i+1)}|) = O(|E^{(\geq i)}|)$. 

Summing over all $i \in [0, L-1]$, we derive that:
\begin{equation}
\label{eq:runtime:1}
\text{The total runtime of  \Cref{alg:uniform-sparsify} is at most } \sum_{i=0}^{L-1} O\left(|E^{(\geq i)}|\right).
\end{equation}
Next,  fix an iteration $i$ of the outer {\sc For} loop in  \Cref{alg:build}, and focus on the call to {\sc Degree-Split}$(E^{(\geq i)} \setminus F^{(i)})$. The inner {\sc While} loop ensures that $\text{deg}_{E^{(\geq i)} \setminus F^{(i)}}(v) > (1/\epsilon)$ for all  $v \in V^{(\geq i+1)}$. By  \Cref{cl:static:deg:split:2},  the degree of every concerned node is (roughly) halved by a call to {\sc Degree-Split}$(.)$. Hence, $|E^{(\geq i+1)}| = O\left((1/2) \cdot |E^{(\geq i)}| \right)$ for all $i \in [0, L-1]$. Plugging this back into~(\ref{eq:runtime:1}),  the lemma follows since $\sum_{i=0}^{L-1} O\left(|E^{(\geq i)}|\right) = O\left( |E^{(\geq 0)}| \right) = O(|E|)$. 
\end{proof}

We will next show that the subgraph $H = (V, F)$ returned by our algorithm is a good matching-sparsifier.
 Towards this end, we first derive the following important  claim.

\begin{claim}(Informal)
\label{cl:informal:node:weight:split}
For all $v \in V$ and  $i \in [0, L-1]$, we have $\gamma^{(i+1)}(v) \simeq (1+\epsilon) \cdot \gamma^{(i)}(v)$. 
\end{claim} 

\begin{proof}
Let us track how starting from $\gamma^{(i)}$,  the weight-function $\gamma^{(i+1)}$ is constructed during round $i$. Recall that $\text{{\sc Support}}\left( \gamma^{(i)} \right) := E^{(\geq i)} \bigcup_{j=0}^{i-1} F^{(j)}$.  During round $i$, we first identify the subset  $F^{(i)} \subseteq E^{(\geq i)}$, and then identify another subset  $E^{(\geq i+1)} \subseteq E^{(\geq i)} \setminus F^{(i)}$. As we switch from $\gamma^{(i)}$ to $\gamma^{(i+1)}$, the following three events occur: (1) The weights of the edges $e \in F^{(i)} \bigcup_{j=0}^{i-1} F^{(j)}$ do not change. (2) The edges $e \in \left( E^{(\geq i)} \setminus F^{(i)} \right) \setminus E^{(\geq i+1)}$ get discarded from the support of $\gamma^{(i+1)}$. (3) The weights of the remaining edges $e \in E^{(\geq i+1)}$ get doubled. 

Now, fix any node $v \in V$. The  claim follows from our analysis of the two cases below.

\smallskip
\noindent {\em Case 1:} $v \in V^{(\geq i+1)}$. In this case, the inner {\sc While} loop in \Cref{alg:build} ensures that $\text{deg}_{E^{(\geq i)} \setminus F^{(i)}}(v) > (1/\epsilon)$. Hence, applying   \Cref{cl:static:deg:split:2}, we get: $\text{deg}_{E^{(\geq i+1)}}(v) \simeq (1\pm \epsilon) \cdot (1/2) \cdot \text{deg}_{E^{(\geq i)} \setminus F^{(i)}}(v)$. To summarize, (about) half the edges in $E^{(\geq i)} \setminus F^{(i)}$ that are incident on $v$ get discarded from the support of $\gamma^{(i+1)}$, while the remaining edges in $E^{(\geq i)} \setminus F^{(i)}$ double their weights. In contrast, the edges in $F^{(i)} \bigcup_{j=0}^{i-1} F^{(j)}$ that are incident on $v$ do not change their weights at all. This implies that $\gamma^{(i+1)}(v) \simeq (1\pm \epsilon) \cdot \gamma^{(i)}(v)$.

\smallskip
\noindent {\em Case 2:} $v \notin V^{(\geq i+1)}$. In this case, every edge $(u, v) \in \text{{\sc Support}}\left( \gamma^{(i)} \right)$ continues to remain in the support of $\gamma^{(i+1)}$ with the same weight.  Hence, we get: $\gamma^{(i+1)}(v) = \gamma^{(i)}(v)$.
\end{proof}

\begin{lemma}(Informal)
\label{cor:informal:node:weight:split}
\label{lm:informal:node:weight:split}
The weight-function $h : F \rightarrow [0, 1]$ satisfies three properties:
\begin{enumerate}
\item $h(e) < \beta$ for all edges $e \in F$.
\item $w(v) \simeq (1\pm \epsilon L) \cdot h(v)$ for all nodes $v \in V$.
\item $size(w) \simeq (1\pm \epsilon L) \cdot size(h)$.
\end{enumerate}
\end{lemma}

\begin{proof}
Before the start of round $0$, we have $\gamma^{(0)}(e) = \lambda$ for all edges $e \in E^{(\geq 0)}$. Subsequently, in each round $i \in [0, L-1]$, the weight of each edge  $e \in E^{(\geq i+1)}$ gets doubled. Hence, we have $h(e) = \gamma^{(L)}(e) \leq 2^{L} \lambda < \beta$ for all $e \in F$. This proves part-(1) of the lemma. 

Next, fix any node $v \in V$. Before the start of round $0$, we have $\gamma^{(0)} = w$ and hence $\gamma^{(0)}(v) = w(v)$. Subsequently, after each round $i \in [0, L-1]$,  \Cref{cl:informal:node:weight:split} guarantees that $\gamma^{(i+1)}(v) \simeq (1\pm \epsilon) \cdot \gamma^{(i)}(v)$. This gives us: $h(v) = \gamma^{(L)}(v) \simeq (1\pm \epsilon)^L \cdot  \gamma^{(0)}(v) \simeq (1 \pm \epsilon L) \cdot \gamma^{(0)}(v) \simeq (1\pm \epsilon L) \cdot w(v)$. Finally, summing this (approximate) equality over all nodes $v \in V$, we get: $size(w) \simeq (1\pm \epsilon L) \cdot size(h)$. This proves part-(2) and part-(3) of the lemma.
\end{proof}

\noindent {\bf Levels of nodes and edges:} The {\em level} of a node $v \in V$ is defined as $\ell(v) := \max \{ i \in [0, L] : v \in V^{(\geq i)} \}$. Similarly, the {\em level} of an edge $e \in E$ is defined as $\ell(e) := \max \{ i \in [0, L] : e \in E^{(\geq i)} \}$. Since $V^{(i)} = V^{(\geq i)} \setminus V^{(\geq i+1)}$ for all $i \in [0, L]$, it follows that $\ell(v) = i$ iff $v \in V^{(i)}$.

\begin{observation}
\label{ob:edge:level}
For all $(u, v) \in F$, we have $\ell(u, v) = \min(\ell(u), \ell(v))$ and $(u, v) \in F^{(\ell(u, v))}$.
\end{observation}

\begin{proof}
Consider any edge $(u, v) \in F = \bigcup_{i=0}^L F^{(i)}$. W.l.o.g., suppose that $(u, v) \in F^{(j)}$ for some $j \in [0, L]$. Before the start of round $0$, we have $(u, v) \in E^{(\geq 0)}$. During each round $i \in [0, j-1]$, the edge $(u, v)$ gets included in the set $E^{(\geq i+1)}$, and both its endpoints $u, v$ get included in the set $V^{(\geq i+1)}$. At round $j$, one of its endpoints (say, $u$) gets included in  $V^{(j)}$, and the edge $(u, v)$ also gets included in $F^{(j)}$. Since $F^{(j)} \subseteq E^{(\geq j+1)} \setminus E^{(\geq j)}$, we infer that $\ell(u) = j$, $\ell(v) \geq j$, and $\ell(u, v) = \max \left\{ i \in [0, L] : (u, v) \in E^{(\geq i)} \right\} = j = \min(\ell(u), \ell(v))$.  
\end{proof}

\begin{observation}
\label{ob:edge:weight}
For every edge $(u, v) \in F$, we have $h(u, v) = 2^{\ell(u, v)} \cdot \lambda$. 
\end{observation}

\begin{proof}
Suppose that $\ell(u, v) = i \in [0, L]$, and hence $(u, v) \in F^{(i)}$. Before the start of round $0$, we have $(u, v) \in E^{(\geq 0)}$ and $\gamma^{(0)}(u, v) = \lambda$. During each round $j \in [0, i-1]$, the edge $(u, v)$ gets included in $E^{(\geq j+1)}$ and we double its weight, i.e., we set $\gamma^{(j+1)}(u,v) := 2 \cdot \gamma^{(j)}(u, v)$. Thus, at the start of round $i$, we have $(u, v) \in E^{(\geq i)}$ and $\gamma^{(i)}(u, v) = 2^i  \cdot \lambda$. During round $i$, the edge $(u,v)$ gets included in the set $F^{(i)}$ and its weight is frozen for the subsequent rounds, so that we get: $2^{i} \cdot \lambda = \gamma^{(i)}(u, v) = \gamma^{(i+1)}(u, v) = \cdots = \gamma^{(L)}(u, v) = h(u, v)$.    
\end{proof}

\begin{lemma}
	\label{lm:static:orientation}
	\label{cor:lm:static:orientation}	
	The graph $H = (V, F)$  has arboricity at most $O\left(\epsilon^{-1} +  \beta^{-1} \right)  = O(\log n)$. 
\end{lemma}	

\begin{proof}
For any  nodes $u, v \in V$ with $\ell(u) = \ell(v) = i < L$, we say that $u$ was assigned its level {\em before} $v$ iff we had $u \in V^{(i)}$ just before the iteration of the inner {\sc While} loop in  \Cref{alg:build} which adds $v$ to $V^{(i)}$.  We now define the following orientation of the graph $H = (V, F)$:
\begin{itemize}
\item Consider any edge $(u, v) \in F$. W.l.o.g.~suppose that $\ell(u) \leq \ell(v)$. If $\ell(u) < \ell(v)$, then the edge   is orientated {\em from} $u$ {\em towards} $v$. Otherwise, if  $\ell(u) = \ell(v) = L$, then the edge  is oriented in any arbitrary direction. Finally, if $\ell(u) = \ell(v) < L$ and (say) the node $u$ was assigned its level before the node $v$, then  the edge  is oriented {\em from} $u$ {\em towards} $v$.
\end{itemize}
Fix any node $x \in V$. Define $\text{Out}_F(x) := \{ (x, y) \in F : \text{the edge } (x, y) \text{ is oriented away from } x\}$. We will show that $|\text{Out}_F(x)| \leq O\left(\epsilon^{-1} + \beta^{-1}\right)$. The lemma will then follow from~(\ref{eq:bound:epsilon:new}).

\smallskip
\noindent {\em (Case 1):} $\ell(x) = i < L$. Let $X^- \subseteq V^{(\geq i)}$ be the set of nodes in $V^{(\geq i)}$ that are assigned the level $i$ {\em before} the node $x$. In  words, the symbol $X^-$ denotes the status of the set $V^{(i)}$ just before $x$ gets added to $V^{(i)}$ in  \Cref{alg:build}.  For every edge $(x, y) \in \text{Out}_F(x)$, we have $y \in V^{\geq i} \setminus X^{-}$ and $(x, y) \in E^{(\geq i)}$. Hence, it follows that $|\text{Out}_F(x)| \leq \text{deg}_{E^{(\geq i)}}(x, V^{(\geq i)} \setminus X^-) \leq \epsilon^{-1}$. 

\smallskip
\noindent {\em (Case 2):} $\ell(x)  = L$. Consider any edge $(x, y) \in \text{Out}_F(x)$. Clearly, this implies that $\ell(y) = L$, and hence $\ell(x, y) = L$ by \Cref{ob:edge:level}. Thus, by Observation~\ref{ob:edge:weight} we have  $h(x, y) = 2^{L} \cdot \lambda \geq \beta/2$. In other words,  $h(x, y) \geq \beta/2$ for all $(x, y) \in \text{Out}_F(x)$.  Now,  part-(2) of  \Cref{cor:informal:node:weight:split} implies that: $w(x) = \Omega(h(x)) = \Omega\left( \sum_{(x, y) \in \text{Out}_F(x)} h(x, y)\right)  = \Omega\left( |\text{Out}_F(x)| \cdot (\beta/2) \right)$.
Accordingly, we get: $w(x) = \Omega\left( |\text{Out}_F(x)| \cdot \beta \right)$, and hence: $|\text{Out}_F(x)| = O\left( \beta^{-1} \cdot w(x) \right) = O(\beta^{-1})$. The last inequality holds since $w(x) \leq 1$.
\end{proof}

To summarize, our static algorithm runs in linear time (\Cref{lm:static:runtime}),  returns a subgraph $H = (V, F)$ with bounded arboricity (\Cref{cor:lm:static:orientation}), and this subgraph $H$ admits a fractional matching that closely approximates the input $\lambda$-uniform matching $w$ in $G$ (\Cref{lm:informal:node:weight:split}).

\section{Dynamically Sparsifying a Uniform Fractional Matching}
\label{sec:dyn:uniform}

In this section, we will present a dynamic algorithm for sparsifying a uniform fractional matching, which will be referred to as {\sc Dynamic-Uniform-Sparsify}$(G = (V, E), \lambda)$.  The input to this algorithm is a dynamic graph $G = (V, E)$ that keeps changing via a sequence of updates (edge insertions/deletions), and a fixed parameter $\delta / n^2 \leq \lambda < \beta$. Throughout this sequence of updates, it is guaranteed that the graph $G$ admits a valid $\lambda$-uniform fractional matching $w$.  We  will show how to maintain a subgraph $H_{(a)} = (V, F_{(a)})$ of this dynamic graph $G = (V, E)$, with $F_{(a)} \subseteq E$, that is a good matching-sparsifier of $G$ with respect to $w$.

Our dynamic algorithm  will be heavily based on the static algorithm from  \Cref{sec:static}. We  now introduce a couple of (informal) terms that relate  to various aspects of  this static algorithm. These terms  will be very useful in the ensuing discussion. First, for each $i \in [0, L]$,  the term {\em level-$i$-structure} will refer to the following sets: $E^{(\geq i)}, V^{(\geq i)}, V^{(i)}$ and $F^{(i)}$.  Second, the term {\em hierarchy} will refer to the union of the  level-$i$-structures over all $i \in [0, L]$. 


We will maintain a partition of the edge-set $E$ into two subsets: $E_{(a)}$ and $E_{(p)}$. The edges in $E_{(a)}$ (resp., $E_{(p)}$) will be called {\em active} (resp., {\em passive}). We will let $G_{(a)} := (V, E_{(a)})$ and $G_{(p)} := (V, E_{(p)})$ respectively denote the active and passive subgraphs of the input graph $G = (V, E)$. Our dynamic algorithm will make a {\em lazy attempt} at mimicking   the static algorithm from \Cref{sec:static}, when the latter receives the active subgraph $G_{(a)}$ as input.

\smallskip
\noindent {\bf Preprocessing:} 
At preprocessing, we set $E_{(p)} := \emptyset$ and $E_{(a)} := E$, and then call   {\sc Static-Uniform-Sparsify}$(G_{(a)} = (V, E_{(a)}), \lambda)$, as described  in  \Cref{alg:uniform-sparsify}. It returns the hierarchy, where for each $i \in [0, L]$ the level-$i$-structure consists of $E^{(\geq i)}, V^{(\geq i)}, F^{(i)}, V^{(i)}$. Finally, for each $i \in [0, L]$, we  initialize  a set $D^{(\geq i)} := \emptyset$. This concludes the preprocessing step.

\smallskip
\noindent {\bf Handling an edge-insertion:} When an edge $e$ gets inserted into  the input graph $G = (V, E)$, we call the subroutine {\sc Handle-Insertion}$(e, \lambda)$, as described in \Cref{alg:handle:insertion}. This classifies the edge $e$ as passive, and sets $E_{(p)} \leftarrow E_{(p)} \cup \{e \}$. If the previous step does not violate Invariant~\ref{inv:insertion}, then we are done. Otherwise, if Invariant~\ref{inv:insertion} gets violated, then we throw away the existing hierarchy and all its associated structures (such as  the sets $D^{(\geq i)}$), and perform the preprocessing step again on the current input graph $G$.

\begin{invariant}
\label{inv:insertion}
 $|E_{(p)}| \leq \epsilon \cdot |E_{(a)}|$. 
\end{invariant}

\noindent {\bf Handling an edge-deletion:} When an edge $e$ gets deleted from $G$, we call the subroutine {\sc Handle-Deletion}$(e, \lambda)$, as described in \Cref{alg:handle:deletion}.  If $e$ was already passive, then it simply gets  removed  from the set $E_{(p)}$, and we are done. Henceforth, we assume that $e$ was active, and at level $\ell(e) = k$, just before getting deleted.\footnote{See the paragraph just before Observation~\ref{ob:edge:level} for the definition of the level of an edge.} 

First, we remove  $e$ from the set $E_{(a)}$, because the edge is no longer present in  $G$. Next, for every $i \in [0, k]$, we  insert  $e$ into the set $D^{(\geq i)}$. From now on, we will refer to $e$ as a {\em dead edge}. Intuitively, the edge $e$, even after getting deleted, continues to be present in the level-$i$-structure for each $i \in [0, k]$. Thus, up until this point, the hierarchy does not change. 

Next, we check if the previous steps lead to a  violation of Invariant~\ref{inv:deletion}. If Invariant~\ref{inv:deletion} continues to remain satisfied, then we are done. Otherwise, we find the minimum index $j \in [0, L]$ such that $\left| D^{(\geq j)} \right| > \epsilon \cdot |E^{(\geq j)}|$, and then perform the following operations: (1) For every $i \in [j, L]$, we delete the dead edges $D^{(\geq i)}$ from the level-$i$-structure and reset  $D^{(\geq i)} \leftarrow \emptyset$. (2) Finally, we call the subroutine {\sc Rebuild}$(j, \lambda)$ as described in  \Cref{alg:build}.

\begin{invariant}
\label{inv:deletion}
$\left| D^{(\geq i)} \right| \leq \epsilon \cdot \left| E^{(\geq i)} \right|$ for all $i \in \left[0, L\right]$. 
\end{invariant}

	\begin{algorithm}[H]
	\caption{{\sc Handle-Insertion}$(e, \lambda)$}
	\label{alg:handle:insertion}
		\begin{algorithmic}
			\State $E_{(p)} \leftarrow E_{(p)} \cup \{e \}$.
                          \If{$\left| E_{(p)} \right| > \epsilon \cdot \left| E_{(a)} \right|$}
                          \State Call the subroutine {\sc Clean-Up}$(0, \lambda)$.
                          \State $E_{(a)} \leftarrow E_{(a)} \cup E_{(p)}$. 
                          \State $E_{(p)} \leftarrow \emptyset$.
                          \State Call the subroutine {\sc Static-Uniform-Sparsify}$\left(G_{(a)} := (V, E_{(a)}), \lambda \right)$.	
                          \EndIf
		\end{algorithmic}
	\end{algorithm}

	\begin{algorithm}[H]
	\caption{{\sc Clean-Up}$(j, \lambda)$}
	\label{alg:clean:up}
		\begin{algorithmic}
			 \For{all $i = j$ to $L$}
                          \State $E^{(\geq i)} \leftarrow E^{(\geq i)} \setminus D^{(\geq i)}$. 
                          \State $F^{(i)} \leftarrow F^{(i)} \setminus D^{(\geq i)}$. 
                          \State $D^{(\geq i)} \leftarrow \emptyset$. 
                          \EndFor		
                  \end{algorithmic}
	\end{algorithm}

	\begin{algorithm}[H]
	\caption{{\sc Handle-Deletion}$(e, \lambda)$}
	\label{alg:handle:deletion}
		\begin{algorithmic}
			\State \If{$e \in E_{(p)}$}
			\State $E_{(p)} \leftarrow E_{(p)} \setminus \{ e \}$.
			\Else
			\State $k \leftarrow \ell(e) :=  \max \left\{i \in [0, L]  : e \in E^{(\geq i)} \right\}$. 
			\State $E_{(a)} \leftarrow E_{(a)} \setminus \{e \}$.
			\For{$i = 0$ to $k$}
			\State $D^{(\geq i)} \leftarrow D^{(\geq i)} \cup \{e\}$.
			\EndFor
                          \If{$\left| D^{(\geq i)} \right| > \epsilon \cdot \left| E^{(\geq i)} \right|$ for some index $i \in \left[0, L \right]$}
                          \State Let $j$ be the minimum index $i \in \left[0, L \right]$ for which $\left| D^{(\geq i)} \right|  > \epsilon \cdot \left| E^{(\geq i)} \right|$. 
                           \State Call the subroutine {\sc Clean-Up}$(j, \lambda)$.
                           \State Call the subroutine {\sc Rebuild}$\left(j, \lambda \right)$.                         
			\EndIf
			\EndIf
			\end{algorithmic}
	\end{algorithm}

Note that $E_{(p)}, E_{(a)}, E^{(\geq i)}, D^{( \geq i)}$,  $F^{(i)}$ represent global variables in these pseudocodes.

\medskip

To summarize, we satisfy Invariant~\ref{inv:insertion} and Invariant~\ref{inv:deletion}, and handle the updates to $G$ in a lazy manner. Newly inserted edges are classified as passive, and they are completely ignored in the hierarchy unless their number becomes sufficiently large compared to the total number of active edges, at which point we rebuild everything from scratch. In contrast, when an active edge gets deleted  from some level $i \in [0, L]$, it is classified as dead and it continues to be present in the level-$j$-structure for all $j \in [0, i]$. Finally, if we notice that for some $k \in [0, L]$  the level-$k$-structure has too many dead edges $D^{(\geq k)}$, then we remove all the dead edges from every level-$j$-structure with $j \in [k, L]$, and  rebuild these structures  from scratch.

From  \Cref{sec:static}, recall that $F := \bigcup_{i=0}^L F^{(i)}$. For any set of edges $E'$, we will use the notation $E'_{(a)} := E' \cap E_{(a)}$ to denote the subset of edges in $E'$ that are active in the current input graph $G = (V, E)$. Accordingly, we define $F_{(a)} := F \cap E_{(a)}$ and $H_{(a)} := (V, F_{(a)})$. 

 \Cref{lm:dyn:uniform:sparsify:1} and  \Cref{lm:dyn:uniform:sparsify:2} below should respectively be seen as analogues of \Cref{cor:lm:static:orientation} and \Cref{lm:informal:node:weight:split} from  \Cref{sec:static}. They show that the subgraph $H_{(a)} = (V, F_{(a)})$ is a good matching sparsifier of the input dynamic graph $G = (V, E)$. Intuitively,  \Cref{lm:dyn:uniform:sparsify:1} and  \Cref{lm:dyn:uniform:sparsify:2} hold because Invariant~\ref{inv:insertion} and Invariant~\ref{inv:deletion} ensure that throughout the sequence of updates, the hierarchy maintained by our dynamic algorithm is {\em very close} to the hierarchy constructed by the  algorithm from  \Cref{sec:static} when it receives the current graph $G = (V, E)$ as input. Due to space constraints, the proofs of these two lemmas are deferred to Appendix~\ref{sec:dyn:uniform:proofs}.

\begin{lemma}
\label{lm:dyn:uniform:sparsify:1}
The graph $H_{(a)} = (V, F_{(a)})$ has arboricity at most  $O(\epsilon^{-1} +  \beta^{-1}) = O(\log n)$.
\end{lemma}

\begin{lemma}
\label{lm:dyn:uniform:sparsify:2}
The graph $H_{(a)}$ admits a fractional matching $h' : F_{(a)} \rightarrow [0, 1]$ such that:
\begin{enumerate}
\item For every edge $e \in F_{(a)}$, we have $h'(e) < \beta$.
\item For every node $v \in V$, we have $h'(v) \leq w(v)$. 
\item We have $size(w) \leq (1 + 60 \epsilon \cdot \log (\beta/\lambda)) \cdot size(h')$.
\end{enumerate}
\end{lemma}

\begin{lemma}
\label{lm:dyn:uniform:update:time}
 The dynamic algorithm {\sc Dynamic-Uniform-Sparsify}$(G, \lambda)$  has an amortized update time of  $O\left(\epsilon^{-1} \cdot \log (\beta/\lambda) \right) = O(\log^2 n)$.
\end{lemma}

\begin{proof}
Define a potential function $\Phi := |E_{(p)}| + \sum_{i = 0}^L \left|D^{(\geq i)} \right|$. Insertion of an edge increases the potential $\Phi$ by at most one unit, as the newly inserted edge gets classified as passive. On the other hand, deletion of an edge $e$ increases the potential $\Phi$ by at most $L+1$ units, since $e$ gets added to each of the sets $D^{(\geq 0)}, \ldots, D^{(\geq \ell(e))}$, and $\ell(e) \leq L$. To summarize, each update in $G$ creates at most $O(L)$ units of new potential. We will show that whenever our dynamic algorithm spends $T$ units of time, the potential $\Phi$ drops by at least $\Omega( \epsilon \cdot T)$. Since $\Phi$ is always  $\geq 0$, this implies the desired  amortized update time of $O(\epsilon^{-1} \cdot L) = O\left(\epsilon^{-1} \cdot \log (\beta/\lambda)\right)$.

Consider the insertion of an edge $e$ into $G = (V, E)$, and suppose that we call  {\sc Static-Uniform-Sparsify}$(G_{(a)}, \lambda)$ while handling this insertion. Let  $m_{(a)} := |E_{(a)}|$, $m_{(p)} := |E_{(p)}|$, $m_{(d)} := |D^{(\geq 0)}|$ and $m := |E|$, just before  $e$ gets inserted. \Cref{inv:insertion} and \Cref{inv:deletion} respectively ensure that $m_{(p)} = \epsilon \cdot m$ and $m_{(d)} \leq \epsilon \cdot m$. By \Cref{lm:static:runtime}, the call to {\sc Static-Uniform-Sparsify}$(G_{(a)}, \lambda)$ takes $O(m)$ time. Thus, the total time to handle this edge insertion is given by $T := O(m + m_{d}) = O(m)$. On the other hand, when our algorithm finishes handling this edge insertion, we have $E_{(p)} = \emptyset$, and hence the potential $\Phi$ decreases by at least $m_{(p)} = \epsilon \cdot m$ units. In other words, the drop in the potential $\Phi$ is at least $\Omega(\epsilon \cdot T)$. 

Next, consider the deletion of an edge $e$ from $G = (V, E)$, and suppose that while handling this deletion we call the subroutine {\sc Rebuild}$(k, \lambda)$ for some $k \in [0, L]$.  Just before $e$ gets deleted, let $m_{(d)}^{(\geq k)} := \left| D^{(\geq k)} \right|$ and $m^{(\geq k)} := E^{(\geq k)}$. \Cref{inv:deletion} ensures that $m_{(d)}^{(\geq k)} = \epsilon \cdot m^{(\geq k)}$. By \Cref{lm:static:runtime}, the call to {\sc Rebuild}$(k, \lambda)$ takes $O\left(m^{(\geq k)} \right)$ time. Hence, excluding the time it takes to identify the level $k$, which is $O(L) = O(\log (\beta/\lambda)) = O(\log n)$ in the worst-case, our dynamic algorithm spends $T = O\left(m^{(\geq k)} \right)$ time to handle this edge deletion. On the other hand, this decreases the potential $\Phi$ by at least $m^{(\geq k)}_{(d)} = \epsilon \cdot m^{(\geq k)}$, since once we are done processing this edge deletion, we have $D^{(\geq k)} = \emptyset$. So the potential $\Phi$ drops by $\Omega(\epsilon \cdot T)$.
\end{proof}

\section{Dynamically Sparsifying an Arbitrary Fractional Matching}
\label{sec:dyn:general}

In this section, we  briefly  sketch   our dynamic algorithm for maintaining a matching-sparsifier as specified by \Cref{th:inf:main}. The full version of the algorithm appears in \Cref{app:sec:dyn:general}.

The input is a dynamic graph $G = (V, E)$ with $n$ nodes, and a ({\em not} necessarily uniform) fractional matching $w : E \rightarrow [0, 1]$ in $G$.  An ``update'' either inserts/deletes an edge in $G$ or changes the weight $w(e)$ of an existing edge $e$ in $G$. Our  algorithm works in three  steps.

\smallskip
\noindent {\bf Step I (Discretizing $w$):}  For every integer $j \geq 0$, define $\lambda_j := (\beta/n^2) \cdot (1+\beta)^j$. Let $K$ be the largest integer $j$ such that $\lambda_j < \beta$. We now discretize $w$ to get a new fractional matching $\hat{w} : E \rightarrow [0, 1]$, which is defined as follows. Consider any edge $e \in E$. If $w(e) < \lambda_0$, then $\hat{w}(e) := 0$. Else if $\lambda_0 \leq w(e) < \beta$, then $\hat{w}(e) := \lambda_i$ where $i$ is the unique integer such that $\lambda_i \leq w(e) < \lambda_{i+1}$. Otherwise, if $w(e) \geq \beta$, then $\hat{w}(e) := w(e)$. 

For each $i \in [0, K]$, let $E_i$ denote the subset of edges $e \in E$ with $\hat{w}(e) = \lambda_i$,  let $G_i := (V, E_i)$, and let $w_i : E_i \rightarrow [0, 1]$ be the restriction of the fractional matching $\hat{w}$ onto the set $E_i$ (i.e., $w_i$ is a $\lambda_i$-uniform fractional matching in $G_i$). Finally, define the subset of edges $E_{\geq \beta} := \{ e \in E : \hat{w}(e) = w(e) \geq \beta\}$, and let $G_{\geq \beta} := (V, E_{\geq \beta})$. In the dynamic setting, we can easily maintain the subgraphs $G_0, \ldots, G_K, G_{\geq \beta}$ of the input graph $G$ {\em on the fly}.

\smallskip
\noindent {\bf Step II (Sparsifying each $G_i$):} For each $i \in [0, K]$, we maintain a sparsifier of $G_i$ with respect to $w_i$, with the help of the dynamic algorithm from \Cref{sec:dyn:uniform}. Specifically, let $H_i = (V, F_i)$ denote the sparsifier $H_{(a)} = (V, F_{(a)})$ maintained by the algorithm {\sc Dynamic-Uniform-Sparsify}$(G_i, \lambda_i)$ from \Cref{sec:dyn:uniform} (thus, we have $F_i \subseteq E_i$). 

\smallskip
\noindent {\bf Step III (Putting everything together):} Let $E_S := \bigcup_{i=0}^K F_i \bigcup E_{\geq \beta}$. In \Cref{app:sec:dyn:general}, we show that the subgraph $S := (V, E_S)$ of $G$ satisfies all the five conditions stated in \Cref{th:inf:main}.

\newpage

\section*{Appendix}
\appendix

\section{Missing Proofs from  \Cref{sec:dyn:uniform}}
\label{sec:dyn:uniform:proofs}

This section presents complete proofs of  \Cref{lm:dyn:uniform:sparsify:1,lm:dyn:uniform:sparsify:2}, and  is organized as follows. 

In \Cref{app:sec:observe}, we present a simple observation and a couple of inequalities that will be very useful later on. In \Cref{app:distance}, we introduce and analyze a distance measure between two weight-functions defined on the same node-set. In~\Cref{sec:analogous:static}, we present an {\em analogous} static algorithm and prove that it outputs a hierarchy which is always the same as the hierarchy maintained by the dynamic algorithm from \Cref{sec:dyn:uniform}. Armed with this observation, from this point onward we only analyze the hierarchy returned by this analogous static algorithm in our proofs. In \Cref{app:basic:property}, we derive some key properties of this analogous static algorithm. Finally, using these properties, we prove \Cref{lm:dyn:uniform:sparsify:2} in \Cref{sec:lm:dyn:uniform:sparsify:2}, and \Cref{lm:dyn:uniform:sparsify:1} in \Cref{sec:lm:dyn:uniform:sparsify:1}.

\medskip

Throughout \Cref{sec:dyn:uniform:proofs},  we write $x = y \pm z$ to denote that $x \in [y-z, y+z]$.

\subsection{Some Basic Observations}
\label{app:sec:observe}

We start with the following simple observation, which shows that the sets of dead-edges present at different levels of the hierarchy form a laminar family.

\begin{observation}
\label{ob:dead:edges:nested}
We always have: $D^{(\geq 0)} \supseteq D^{(\geq 1)} \supseteq \cdots \supseteq D^{(\geq L-1)} \supseteq D^{(\geq L)}$.
\end{observation}

\begin{proof} Clearly, the observation holds immediately after the preprocessing step. 

By inductive hypothesis, suppose that the observation holds just before a given update in $G$. While handling the current update in $G$, the sets $D^{(\geq i)}$ can change only in one of the following two manners. (1) Some edge $e$ at a level $\ell(e) = k \in [0, L]$ gets marked as {\em dead}, and gets added to each of the sets $D^{(\geq 0)}, \cdots, D^{(\geq k)}$. (2) We identify some level $i \in [0, L]$, reset $D^{(\geq j)} \leftarrow \emptyset$ for all $j \in [i, L]$, and then rebuild the level-$j$-structure for all $j \in [i, L]$. 

Neither step (1) nor step (2) above violates the observation. Hence, the observation continues to remain valid even after our dynamic algorithm handles the current update.
\end{proof}

Since $\lambda \geq \delta / n^2$, it follows that $L \leq \log (\beta / \lambda) \leq 2 \cdot \log n$. We will use this upper bound on $L$ throughout the rest of this section. Now, from~(\ref{eq:bound:epsilon:new}) we can infer that:

\begin{eqnarray}
	0 < \epsilon \leq \min \left(10^{-4}, \frac{1}{500 \cdot \log (\beta /\lambda)} \right). \label{eq:bound:epsilon} 
\end{eqnarray}

Finally, we state a simple claim that will be very useful in our subsequent calculations.

\begin{claim}
\label{cl:tool:exponential}
For all $i \geq 1$ and  $0 \leq z \leq 1/(2 i)$, we have: 
$$(1 + z)^i \leq  (1 + 2  i z ) \text{ and } (1-z)^i \geq (1-2 i z).$$
\end{claim}

\subsection{A Measure of Distance between Two Weight-Functions}
\label{app:distance}

\Cref{def:dist} introduces a measure of distance between two weight-functions that are defined on the same node-set, but not necessarily on the same edge-set. \Cref{prop:triangle} captures the fact that this distance function  satisfies triangle inequality. \Cref{prop:triangle:2} lower bounds  the distance between two weight-functions $w'$ and $w''$ in terms of the absolute difference between their respective sizes. \Cref{prop:triangle:3} shows that the lower bound in \Cref{prop:triangle:2} is tight iff one of the weight-functions dominate the other on every node. Finally, \Cref{lm:weight:function:scaling} implies that if we have two weight-functions $w'$ and $w''$ that are sufficiently close to each other (i.e., the distance between them is small), then we can obtain a new weight-function $\tilde{w}$ by {\em scaling down} $w'$ in an appropriate manner, so that the following properties are satisfied: (1) $\tilde{w}$ is dominated by $w'$ on every edge, (2) $\tilde{w}$ is dominated by $w''$ on every node, and (3) $\tilde{w}$ is sufficiently close to $w''$. We will use this lemma in a crucial manner in subsequent sections.

\begin{definition}
	\label{def:dist}
	Consider any two  graphs $G' = (V, E')$ and $G'' = (V, E'')$ with the same node-set $V$, and two weight functions $w' : E' \rightarrow [0, 1]$ and $w'' : E'' \rightarrow [0, 1]$. Then the {\em distance} between $w'$ and $w''$ is defined as: $dist_V(w', w'') = \sum_{v \in V} | w'(v) - w''(v)|$.
\end{definition}

\begin{property}(Triangle Inequality)
	\label{prop:triangle}
	Consider any three graphs $G' = (V, E')$, $G'' = (V, E'')$, $G''' = (V, E''')$, and weight-functions $w' : E' \rightarrow [0, 1]$, $w'' : E'' \rightarrow [0, 1]$, $w''' : E''' \rightarrow [0, 1]$. Then we have $dist_V(w', w''') \leq dist_V(w', w'') + dist_V(w'', w''')$. 
\end{property}

\begin{proof}
	We observe that:
	\begin{eqnarray*}
		dist_V(w', w''') & = & \sum_{v \in V} | w'(v) - w'''(v)| = \sum_{v \in V} \left| \left(w'(v) - w''(v) \right) + \left( w''(v) - w'''(v) \right) \right| \\
		& \leq & \sum_{v \in V} \left| w'(v) - w''(v) \right| +  \sum_{v \in V} \left| w''(v) - w'''(v) \right| \\
		& \leq & dist_V(w', w'') + dist_V(w'', w''').
	\end{eqnarray*}
\end{proof}

\begin{property}
	\label{prop:triangle:2}
	Consider any two graphs $G' = (V, E')$, $G'' = (V, E'')$, and weight-functions $w' : E' \rightarrow [0, 1]$, $w'' : E'' \rightarrow [0, 1]$. Then we have $dist_V(w', w'') \geq 2 \cdot | size(w') - size(w'')|$.
\end{property}

\begin{proof}
	We observe that: 
	\begin{eqnarray*}
		dist_V(w', w'') = \sum_{v \in V} | w'(v) - w''(v)| \geq \left| \sum_{v \in V} w'(v) - \sum_{v \in V} w''(v) \right| = 2 \cdot \left| size(w') -  size(w'') \right|.
	\end{eqnarray*}
\end{proof}

\begin{property}
	\label{prop:triangle:3}
	Consider any two graphs $G' = (V, E')$, $G'' = (V, E'')$, and weight-functions $w' : E' \rightarrow [0, 1]$, $w'' : E'' \rightarrow [0, 1]$ such that: $w'(v) \leq w''(v)$ for all nodes $v \in V$. Then we have: $dist_V(w', w'') = 2 \cdot \left( size(w'') - size(w') \right)$. 
\end{property}

\begin{proof}
	The property holds because:
	\begin{eqnarray*}
		dist_V(w', w'')  & = &   \sum_{v \in V} \left| w'(v) - w''(v) \right|  =  \sum_{v \in V} \left( w''(v) - w'(v) \right) \\
		& = & \sum_{v \in V} w''(v) - \sum_{v \in V} w'(v) \\
		& = & 2 \cdot size(w'') - 2 \cdot size(w').
	\end{eqnarray*}
\end{proof}

\begin{lemma}
	\label{lm:weight:function:scaling}
	Consider any two graphs $G' = (V, E')$, $G'' = (V, E'')$ and weight-functions $w' : E' : \rightarrow [0, 1]$, $w'' : E'' \rightarrow [0, 1]$ such that: $dist_V(w', w'') \leq \alpha \cdot size(w'')$, for $0 \leq \alpha < 2/3$. Then there exists a weight-function $\tilde{w} : E' \rightarrow [0, 1]$ which satisfy the following conditions.
	\begin{enumerate}
		\item $\tilde{w}(e) \leq w'(e)$ for all edges $e \in E'$.
		\item $\tilde{w}(v) \leq w''(v)$ for all nodes $v \in V$.
		\item $size(w'') \leq (1- 3\alpha/2)^{-1} \cdot size(\tilde{w})$.  
	\end{enumerate}
\end{lemma}

\newcommand{\w}{\tilde{w}}

\subsubsection{Proof of  \Cref{lm:weight:function:scaling}}
We  define a weight-function $\w : E' \rightarrow [0, 1]$, where:
\begin{eqnarray}
	\w(u, v)  = \frac{w'(u, v)}{\max\left(1 , \frac{w'(u)}{w''(u)}, \frac{w'(v)}{w''(v)} \right)}  \text{ for every edge } (u, v) \in E'. \label{eq:new:weight:function}
\end{eqnarray}
It immediately follows that $\w(e) \leq w'(e)$ for all $e \in E'$. Thus, the weight-function $\w$ satisfies condition-(1) in the statement of the lemma. Next, for any node $v \in V$, we observe that:
\begin{eqnarray*}
	\w(v) & = & \sum_{(u, v) \in E'}  \frac{w'(u, v)}{\max\left(1 , \frac{w'(u)}{w''(u)}, \frac{w'(v)}{w''(v)} \right)}  \\
	& \leq & \sum_{(u, v) \in E'}  \frac{w'(u, v)}{\frac{w'(v)}{w''(v)}}  = \frac{w''(v)}{w'(v)}  \cdot \sum_{(u, v) \in E'} w'(u, v) = \frac{w''(v)}{w'(v)}  \cdot w'(v) = w''(v).
\end{eqnarray*}
In other words,  $\w$ satisfies condition-(2) in the statement of the lemma. Henceforth, we focus on proving that  $\w$ satisfies the remaining condition-(3) in the statement of the lemma.

For every node $v \in V$, define the set: 
$$Dom(v) = \left\{ (u, v) \in E' : \frac{w'(v)}{w''(v)} \geq \max\left(1, \frac{w'(u)}{w''(u)}\right) \right\}.$$
We say that an edge $(u, v) \in Dom(v)$ is {\em dominated} by its endpoint $v \in V$. Let $Dom(E') := \{ (u, v) \in E' : \text{either } (u, v) \in Dom(u) \text{ or } (u, v) \in Dom(v) \}$ denote the set of edges in $G'$ that are  dominated by at least one endpoint. We now make the following simple observations.
\begin{eqnarray}
	w'(e) - \w(e) & \geq & 0 \text{ for all } e \in Dom(E'); \label{eq:dominated:0} \\
	w'(e) - \w(e) & = & 0 \text{ for all } e \in E' \setminus Dom(E'); \label{eq:dominated:1} \\
	\w(u, v) & = & w'(u, v) \cdot \frac{w''(v)}{w'(v)} \text{ for all } v \in V \text{ and } (u, v) \in Dom(v); \label{eq:dominated:2} \\
	1 - \frac{w''(v)}{w'(v)} & \geq & 0 \text{ for all } v \in V \text{ with } Dom(v) \neq \emptyset. \label{eq:dominated:3}
\end{eqnarray}

\begin{claim}
	\label{cl:weight:scaling:1}
	$\sum_{e \in Dom(v)} \left( w'(e) - \w(e) \right) \leq w'(v) - w''(v)$, for all $v \in V$ with $Dom(v) \neq \emptyset$.
\end{claim}

\begin{proof}
	Fix any node $v \in V$ with $Dom(v) \neq \emptyset$. We prove the claim as follows.
	\begin{eqnarray}
		\sum_{e \in Dom(v)} \left( w'(e) - \w(e) \right) & = & \sum_{e \in Dom(v)} \left( w'(e) - w'(e) \cdot \frac{w''(v)}{w'(v)} \right) \qquad \qquad (\text{follows from~(\ref{eq:dominated:2})}) \nonumber \\
		& = & \sum_{(u, v) \in Dom(v)} w'(u, v) \cdot \left( 1 - \frac{w''(v)}{w'(v)} \right) \nonumber \\
		& \leq & \sum_{(u, v) \in E'} w'(u, v) \cdot \left( 1 - \frac{w''(v)}{w'(v)} \right) \qquad \qquad \ \ \ \   (\text{follows from~(\ref{eq:dominated:3})}) \nonumber \\
		& = & w'(v)  \cdot \left( 1 - \frac{w''(v)}{w'(v)} \right) \nonumber \\
		& = & w'(v) - w''(v). \nonumber
	\end{eqnarray}
\end{proof}
We will now show that $size(\w)$ is very close to $size(w')$. Towards this end, observe that:
\begin{eqnarray*}
	size(w') - size(\w) & = & \sum_{e \in E'} \left(w'(e) - \w(e)\right) \nonumber  \\
	& = & \sum_{e \in Dom(E')} \left( w'(e) - \w(e) \right) + \sum_{e \in E \setminus Dom(E')} \left( w'(e) - \w(e) \right) \nonumber \\
	& = & \sum_{e \in Dom(E')} \left( w'(e) - \w(e) \right)  \qquad \qquad \qquad  \qquad (\text{follows from~(\ref{eq:dominated:1})}) \\
	& \leq &  \sum_{v \in V} \sum_{e \in Dom(v)} \left( w'(e) - \w(e) \right) \qquad \qquad \qquad (\text{follows from~(\ref{eq:dominated:0})})  \\
	& \leq & \sum_{v \in V} \left( w'(v) - \w(v) \right) \qquad \qquad \qquad \qquad (\text{follows from  \Cref{cl:weight:scaling:1}}) \\
	& \leq & \sum_{v \in V} \left| w'(v) - \w(v) \right| \nonumber \\
	& = & dist_V(w', w'') \qquad \qquad \qquad \qquad (\text{follows from Definition~\ref{def:dist}}) \\
	& \leq & \alpha \cdot size(w'').
\end{eqnarray*}
Rearranging the terms in the above inequality, we get:
\begin{equation}
	size(w') \leq  size(\w) + \alpha \cdot size(w''). \label{eq:dominated:6}
\end{equation}
Next,  \Cref{prop:triangle:2} implies that:
\begin{eqnarray*}
	size(w'') - size(w') \leq |size(w'') - size(w') | \leq (1/2) \cdot dist_V(w', w'') \leq (\alpha/2) \cdot size(w'').
\end{eqnarray*}
Rearranging the terms in the above inequality, we get:
\begin{equation}
	\label{eq:dominated:7}
	size(w') \geq \left( 1 - \frac{\alpha}{2} \right) \cdot size(w'').
\end{equation}
From~(\ref{eq:dominated:6}) and~(\ref{eq:dominated:7}), we get:
$$\left( 1 - \frac{\alpha}{2} \right) \cdot size(w'') \leq size(\w) + \alpha \cdot size(w''), \text{ or equivalently, } size(w'') \leq \left( 1 - \frac{3 \alpha}{2} \right)^{-1}  size(\w).$$
Thus, the weight-function $\w$ satisfies condition-(3) in the statement of the lemma as well. This concludes the proof of  \Cref{lm:weight:function:scaling}.

\subsection{An Analogous Static Setting}
\label{sec:analogous:static}

In \Cref{sec:dyn:uniform}, we explained that our dynamic algorithm for sparsifying a uniform fractional matching attempts to mimic the static algorithm from \Cref{sec:static} in a lazy manner. We now place this intuitive connection between these two  algorithms on a formal footing. Specifically, we describe an {\em analogous static algorithm} that takes the following three things as input: 
\begin{itemize}
\item (1) The node-set $V$. 
\item (2) The set of active edges $E_{(a)}$. 
\item (3) A laminar family of sets of dead-edges $D^{(\geq 0)} \supseteq D^{(\geq 1)} \supseteq \cdots \supseteq D^{(\geq L)}$. 

(We can assume that these sets form a laminar family because of \Cref{ob:dead:edges:nested}.)
\end{itemize}
Based on these three inputs, the analogous static algorithm constructs a hierarchy. We present the analogous static algorithm in \Cref{sec:analogous:static:algo}. Subsequently, in \Cref{sec:equivalence}, we explain that the hierarchy returned by our analogous static algorithm precisely coincides with the hierarchy maintained by the dynamic algorithm from \Cref{sec:dyn:uniform}. 

Henceforth, we will heavily exploit this equivalence between the dynamic setting and the analogous static setting. To be more specific, from \Cref{app:basic:property} onward, we will only focus on the hierarchy returned by the analogous static algorithm in our analysis. Everything we derive about this {\em analogous static hierarchy} will hold for the actual hierarchy maintained by the dynamic algorithm, because these two hierarchies are essentially the same.

\subsubsection{The Analogous Static Algorithm}
\label{sec:analogous:static:algo}

For notational convenience, define $D^{(\geq -1)} := D^{(\geq 0)}$ and $M^{(0)} := D^{(\geq -1)} \setminus D^{(\geq 0)} = \emptyset$. 

\medskip
We set $E^{(\geq 0)} := E_{(a)} \cup D^{(\geq 0)}$, $V^{(\geq 0)} := V$ and $E^{(\geq 0)}_* := E^{(\geq 0)}$. We also  define $\gamma^{(0)}_*$ to be a $\lambda$-uniform weight-function with $\text{{\sc Support}}\left( \gamma^{(0)}_* \right) = E^{(\geq 0)}_*$. 

\medskip
Next, we perform the following operations.

\begin{itemize}
	\item {\sc For} $i = 0$ to $L-1$:
	\begin{itemize}
		\item Call the subroutine {\sc Process-Round}$(i, \lambda)$ described below to implement round $i$.
	\end{itemize}
	\item Define $M^{(L)} := E^{(\geq L)}_* \cap \left( D^{(\geq L-1)} \setminus D^{(\geq L)}\right)$. 
	\item Define $E^{(\geq L)} := E^{(\geq L)}_* \setminus M^{(L)}$. 
	\item Define $V^{(L)} := V^{(\geq L)}$ and $F^{(L)} := E^{(\geq L)}$.
\end{itemize}

\medskip
\noindent
This concludes  the algorithm. It now remains to explain how we implement a given round $i$. 

\medskip
\noindent {\bf  The subroutine {\sc Process-Round}$(i, \lambda)$:} At the start of the call to this subroutine,  we have the weight-function $\gamma^{(i)}_*$ defined over $\text{{\sc Support}}\left( \gamma^{(i)}_* \right) = E^{(\geq i)}_* \bigcup_{j = 0}^{i-1} \left(F^{(j)} \cup M^{(j)} \right)$. Restricted to the subset of edges $E^{(\geq i)}_*$, the weight-function $\gamma^{(i)}_*$ is a $2^i \lambda$-uniform. Specifically, this means that $\gamma^{(i)}_*(e) = 2^i \lambda$ for all $e \in E^{(\geq i)}_*$. We also have the set of nodes $V^{(\geq i)}$. Every edge $e \in E^{(\geq i)}_*$ has both its endpoints in $V^{(\geq i)}$. We implement round $i$ as follows. 

First, we identify the set of edges $M^{(i)} := E^{(\geq i)}_* \cap \left( D^{(\geq i-1)} \setminus D^{(\geq i)}\right)$ that no longer participate in the hierarchy from round $i$ onward. We will refer to the edges $e \in E^{(\geq i)}_*$ as the ones that go {\em missing} in round $i$. Define $E^{(\geq i)} := E^{(\geq i)}_* \setminus M^{(i)}$ and $G^{(\geq i)} := \left( V^{(\geq i)}, E^{(\geq i)} \right)$. Next, we iteratively remove  low-degree nodes from $G^{(\geq i)}$ as per \Cref{alg:build:dyn} below.

\begin{algorithm}[H]
	\caption{}
	\label{alg:build:dyn}
	\begin{algorithmic}
		\State $V^{(i)} \leftarrow \emptyset$.
		\While{there is some node $v \in V^{(\geq i)}\setminus V^{(i)}$ with $\text{deg}_{E^{(\geq i)}}\left(v, V^{(\geq i)} \setminus V^{(i)} \right) \leq (1/\epsilon)$}:
		\State $V^{(i)} \leftarrow V^{(i)} \cup \{v\}$.
		\EndWhile
		\State $V^{(\geq i+1)} \leftarrow V^{(\geq i)} \setminus V^{(i)}$. 
		\State $F^{(i)} \leftarrow \{ (u, v) \in E^{(\geq i)} :  \text{ either } u \in V^{(i)} \text{ or } v \in V^{(i)} \}$. 
	\end{algorithmic}
\end{algorithm}

\noindent We next compute $E^{(\geq i+1)}_* \leftarrow \text{{\sc Degree-Split}}\left(E^{(\geq i)} \setminus F^{(i)}\right)$.	Finally, we define the weight-function $\gamma^{(i+1)}_*$ with $\text{{\sc Support}}\left( \gamma^{(i+1)}_* \right) = E^{(\geq i+1)}_* \bigcup_{j = 0}^{i} \left( F^{(j)} \cup M^{(j)} \right)$ as follows.
\begin{equation}
	\gamma^{(i+1)}_*(e) = \begin{cases} \gamma^{(i)}_*(e) & \text{ if } e \in \bigcup_{j=0}^i \left( F^{(j)} \cup M^{(j)} \right); \\
		2 \cdot \gamma^{(i)}_*(e) & \text{ else if } e \in E^{(\geq i+1)}_*.
	\end{cases}
\end{equation}
Thus, restricted to the subset of edges $E^{(\geq i+1)}_*$, the weight-function $\gamma^{(i+1)}_*$ is $2^{i+1} \lambda$-uniform. This concludes the call to the subroutine {\sc Process-Round}$(i, \lambda)$, which implements round $i$.

\subsubsection{Equivalence Between the Analogous Static Algorithm from \Cref{sec:analogous:static:algo} and the Dynamic Algorithm from \Cref{sec:dyn:uniform}}
\label{sec:equivalence}

The main idea behind proving this equivalence is to focus our attention on the sets of missing edges $M^{(i)}$, for all $i \in [0, L]$. To be a bit more specific, the key observation is that if $M^{(i)} = \emptyset$ for some $i \in [0,L]$, then we have $E^{(\geq i)} = E^{(\geq i)}_*$ in that round $i$.

To elaborate on this in more details, note that at preprocessing, we have $D^{(\geq 0)} = D^{(\geq 1)} = \cdots = D^{(\geq L)} = \emptyset$, and $M^{(\geq 0)} = M^{(\geq 1)} = \cdots = M^{(\geq L)} = \emptyset$. At this point in time, it is trivially true that the hierarchy returned by the analogous static algorithm (henceforth referred to as the analogous static hierarchy) coincides with the hierarchy maintained by the dynamic algorithm (henceforth referred to as the dynamic hierarchy).

By inductive hypothesis, suppose that the analogous static hierarchy coincides with the dynamic hierarchy just before the update at a given time-step (say) $\tau$. Now, this update at time $\tau$ can be one of two types, as explained below.

\medskip
\noindent {\bf Type (I): ``Regular'' Update.} No missing edge gets created or destroyed due to such an update at any round. During such an update only one of the following events can occur. (1) The update happens because of an edge insertion, and the newly inserted edge (say) $e$ got classified as passive and was added to the set $E_{(p)}$. In this case, neither the analogous static hierarchy nor the dynamic hierarchy experiences any change due to this update, because the newly inserted edge $e$ does not participate in any of these two hierarchies. (2) The update happens because of the deletion of an edge (say) $e$ which was at the same level (say) $k \in [0, L]$ in both the two hierarchies (by induction hypothesis). In this case, in both the two hierarchies, the edge $e$ gets added to the sets $D^{(\geq 0)}, \ldots, D^{(\geq k)}$, and nothing else changes.

To summarize, by looking back at these two possible cases, we infer that our inductive hypothesis (that the analogous static hierarchy coincides with the dynamic hierarchy) continues to hold after a ``regular'' update.

\medskip
\noindent {\bf Type (II): ``Triggering" Update.} Such an update can occur because of one of the following two reasons. (1) An edge $e$ gets inserted, gets classified as passive, and this leads to a violation of \Cref{inv:insertion}. In this event, we {\em throw away} both the  analogous static hierarchy and the dynamic hierarchy, discard all the dead-edges from our consideration, reclassify all the passive edges as active, and then rebuild both the analogous static hierarchy and the dynamic hierarchy from scratch by performing exactly the same steps (because of the absence of missing edges at every round). (2) An edge $e$ gets deleted, which leads to a violation of \Cref{inv:deletion}. In this event, we identify a certain level $k \in [0, L]$ and discard all the dead-edges at levels $\geq k$ from our consideration (which means that there will no longer be any missing edge at a round $\geq k$ in the analogous static hierarchy). In the dynamic hierarchy, we rebuild the level-$j$ structure for every level $j \in [k, L]$ from scratch. Because there is no longer any missing edge at round $\geq k$ in the analogous static hierarchy, it is easy to verify that if run the analogous static algorithm on this new input, then the hierarchy it will return will continue to coincide with the dynamic hierarchy.

To summarize, by looking back at these two possible cases, we again infer that our inductive hypothesis continues to hold even after a ``triggering'' update.

\subsection{Some Basic Properties of the  Analogous Static Algorithm}
\label{app:basic:property}

Recall the analogous static algorithm from~\Cref{sec:analogous:static:algo}. We define the weight-function $\gamma_{(a)}$ to be the restriction of $\gamma_{*}^{(L)}$ on the active edges, so that $\text{{\sc Support}}\left( \gamma_{(a)} \right) = \text{{\sc Support}}\left( \gamma^{(L)}_* \right) \setminus D^{(\geq 0)}$. Furthermore, for all $e \in \text{{\sc Support}}\left( \gamma_{(a)} \right)$, we have $\gamma_{(a)}(e) = \gamma^{(L)}_*(e)$. Similarly, we define the weight-function $w_{(a)}$ to be the restriction of $w$ on the active edges. Thus,  $w_{(a)}$ is a $\lambda$-uniform weight-function with support $E_{(a)}$, and we have  $w_a(e) = w(e) = \lambda$ for all $e \in E_{(a)}$.

This section is organized as follows. In \Cref{sec:1}, we derive some basic properties of the weight-functions $\gamma^{(i)}_*$ for all $i \in [0, L]$. Next, in \Cref{sec:2}, we  derive some basic properties of the weight-functions $\gamma_{(a)}$ and  $w_{(a)}$. In \Cref{sec:3}, we show that the weight-function $w_{(a)}$ is {\em very close} to the weight-function $\gamma^{(0)}_*$, in the sense that the distance between $w_{(a)}$ and $\gamma^{(0)}_*$ is very small (according to the distance measure introduced in \Cref{app:distance}). Next, in \Cref{sec:4}, we show that the weight-function $\gamma^{(0)}_*$ is very close to the weight-function $\gamma^{(i)}_*$, for all $i \in [1, L]$. Finally, in \Cref{sec:5}, we prove that the weight-function $\gamma^{(L)}_*$ is very close to the weight-function $\gamma_{(a)}$. At this point, by applying triangle inequality (see \Cref{prop:triangle}), we can infer that the weight-function $w_{(a)}$ is very close to the weight-function $\gamma_{(a)}$. We   use this fact in \Cref{sec:lm:dyn:uniform:sparsify:2}, where we prove \Cref{lm:dyn:uniform:sparsify:2}. 

\subsubsection{Some  Observations on the Weight-Functions $\gamma^{(i)}_*$}
\label{sec:1}

For every edge $e \in E^{(\geq 0)}_* = E^{(\geq 0)}$, define its pseduo-level $\ell_*(e) := \max \left\{ i \in [0, L] : e \in E^{(\geq i)}_* \right\}$. 

\begin{observation}
	\label{ob:pseudo:level:weight}
	Consider any $i \in [0, L]$ and any edge $e \in \text{{\sc Support}}\left( \gamma^{(i)}_* \right)$. We have:
	$$
	\gamma^{(i)}_* = \begin{cases} 2^i \cdot \lambda & \text{ if } \ell_*(e) \geq i; \\
		2^{\ell_*(e)} \cdot \lambda & \text{ else if } \ell_*(e) < i.
	\end{cases}
	$$
\end{observation}

\begin{proof}
	We prove this by  induction on $i$. Clearly, the observation holds for $i = 0$. By inductive hypothesis, suppose that the observation holds for some $i < L$, and focus on the implementation of round $i$ as described in \Cref{sec:analogous:static:algo}. During round $i$, we compute the set  $E^{(\geq i+1)}_* \subseteq E^{(\geq i)}_*$ and construct the weight-function $\gamma^{(i+1)}_*$ from  $\gamma^{(i)}_*$. Fix any edge $e \in \text{{\sc Support}}\left( \gamma^{(i)}_* \right)$, and consider the following mutually exclusive and exhaustive cases.
	
	\medskip
	\noindent {\bf Case 1:} $e  \notin E^{(\geq i)}_*$. In this case, we have $\ell_*(e) < i$. The edge $e$ continues to remain in the support of $\gamma^{(i+1)}_*$ and its weight does not change, i.e., we have $\gamma^{(i+1)}_*(e) = \gamma^{(i)}_*(e) = 2^{\ell_*(e)} \cdot \lambda$, where the last equality follows from inductive hypothesis.
	
	\medskip
	\noindent {\bf Case 2:} $e \in E^{(\geq i+1)}_*$. In this case, the weight of $e$ is doubled as we switch from $\gamma^{(i)}_*$ to $\gamma^{(i+1)}_*$. This means that $\gamma^{(i+1)}_*(e) = 2 
	\cdot \gamma^{(i)}_*(e) = 2^{i+1} \cdot \lambda$, where the last equality follows from the inductive hypothesis and the fact that $E^{(\geq i+1)}_* \subseteq E^{(\geq i)}_*$. 
	
	\medskip
	\noindent {\bf Case 3:} $e \in E^{(\geq i)}_* \setminus E^{(\geq i+1)}_*$. In this case, we have $\ell_*(e) = i$. If this edge $e$ continues to remain in the support of $\gamma^{(i+1)}_*$, then its weight does not change and we get: $\gamma^{(i+1)}_*(e) = \gamma^{(i)}_*(e) = 2^i \lambda = 2^{\ell_*(e)} \cdot \lambda$, where the second last equality follows from the inductive hypothesis.

	\medskip
	Thus, we infer that the observation continues to hold even for round $(i+1)$. This proves the inductive step, and concludes the proof of the observation.
\end{proof}

\begin{corollary}
	\label{ob:gamma:L:level}
	For every edge $e \in \text{{\sc Support}}\left( \gamma^{(L)}_* \right)$, we have $\gamma^{(L)}_*(e) = 2^{\ell_*(e)} \lambda$.
\end{corollary}

\begin{proof}
	Since $\ell_*(e) \leq L$, the corollary follows from \Cref{ob:pseudo:level:weight}.
\end{proof}


\begin{corollary}
	\label{ob:gamma:new:1}
	For every $i \in [0, L]$, we have $size\left(\gamma^{(i)}_*\right) \geq 2^i \lambda \cdot \left| E^{(\geq i)} \right|$. 
\end{corollary}

\begin{proof}
	Note that $E^{(\geq i)}_* \subseteq \text{{\sc Support}}\left( \gamma^{(i)}_* \right)$. By \Cref{ob:pseudo:level:weight}, we have $\gamma^{(i)}_*(e) = 2^i \lambda$ for all  $e \in E^{(\geq i)}_*$. Since $E^{(\geq i)} \subseteq E^{(\geq i)}_*$, we get:
	$size\left(\gamma^{(i)}_*\right) \geq \sum_{e \in E^{(\geq i)}} \gamma^{(i)}_*(e) = 2^{i} \lambda \cdot \left| E^{(\geq i)} \right|.$
\end{proof}


\begin{observation}
	\label{ob:gamma*}
	Consider any edge $e \in \text{{\sc Support}}\left( \gamma^{(L)}_* \right) \cap D^{(\geq 0)}$ with $\ell_*(e) = i \in [0, L]$. Then we must have $e \in D^{(\geq i-1)}$.
\end{observation}

\begin{proof}
	Since $\ell_*(e) = i$, we have $e \in E^{(\geq i)}_*$.  By induction on $i$, it is easy to prove that $E^{(\geq i)}_* \cap D^{(\geq 0)} \subseteq D^{(\geq i-1)}$. In words, if a  dead-edge is present in $E^{(\geq i-1)}_*$ at the start of round $i$, then it belongs to $D^{(\geq i-1)}$.  Accordingly, it follows that $e \in D^{(\geq i-1)}$.
\end{proof}


\begin{observation}
	\label{ob:gamma:0:weight}
	$\gamma^{(0)}_*(v) \leq 1$ for all nodes $v \in V$.
\end{observation}

\begin{proof}
	Fix any node $v \in V$. Suppose that we are looking at the snapshot of our dynamic algorithm at current time-step $\tau$. Let $\tau' < \tau$ be the last time-step before $\tau$ at which either one of the following  events occurred: (1) we called the subroutine {\sc Static-Uniform-Sparsify}$(G_{(a)}, \lambda)$ while handing an insertion at time-step $\tau'$ (see \Cref{alg:handle:insertion}), or (2) we called the subroutine {\sc Rebuild}$(0, \lambda)$ while handing a deletion at time-step $\tau'$ (see \Cref{alg:handle:deletion}). 
	
	Immediately after time-step $\tau'$, we had $D^{(\geq 0)} = \emptyset$ and $E^{(\geq 0)}_* = E^{(\geq 0)} = E_{(a)} \cup D^{(\geq 0)} = E$, and hence $\gamma^{(0)}_*(v) = \left| E^{\geq 0}_* \right| \cdot \lambda = |E| \cdot \lambda = w(v) \leq 1$. During the time-interval $(\tau', \tau)$, whenever an edge gets deleted from $G = (V, E)$, we move it from the set $E_{(a)}$ to the set $D^{(\geq 0)}$. Further, whenever an edge gets inserted into $G = (V, E)$, we do {\em not} include it in either of the sets $\left\{ E_{(a)}, D^{(\geq 0)} \right\}$. Accordingly,  the set $E^{(\geq 0)}_* = E_{(a)} \cup D^{(\geq 0)}$ does not change during the time-interval $(\tau', \tau)$. This, in turn, implies that the value of $\gamma^{(0)}_*(v)$ also does not change during the time-interval $(\tau', \tau)$. Hence, even at the current time-step $\tau$, we have $\gamma^{(0)}_*(v) \leq 1$. 
\end{proof}

\subsubsection{The Weight-Functions $\gamma_{(a)}, w_{(a)}$ and Their Basic Properties}
\label{sec:2}

Recall that $\gamma_{(a)}$ is the restriction of $\gamma_{*}^{(L)}$ on the active edges, so that $\text{{\sc Support}}\left( \gamma_{(a)} \right) = \text{{\sc Support}}\left( \gamma^{(L)}_* \right) \setminus D^{(\geq 0)}$. Furthermore, for all $e \in \text{{\sc Support}}\left( \gamma_{(a)} \right)$, we have $\gamma_{(a)}(e) = \gamma^{(L)}_*(e)$. 

From \Cref{sec:dyn:uniform}, recall that $F := \bigcup_{i=0}^L F^{(i)}$ and $F_{(a)} := F \cap E_{(a)}$.

\begin{observation}
	\label{ob:gamma:a:support}
	$\text{{\sc Support}}\left( \gamma_{(a)} \right) = F_{(a)}$. 
\end{observation}

\begin{proof}
	Note that $\text{{\sc Support}}\left( \gamma^{(L)}_* \right) = \bigcup_{i=0}^L \left(F^{(i)} \cup M^{(i)}\right)$. Since $\bigcup_{i=0}^L M^{(i)} \subseteq D^{(\geq 0)}$, we get:
	\begin{eqnarray*}
		\text{{\sc Support}}\left( \gamma_{(a)} \right) & = & \text{{\sc Support}}\left( \gamma^{(L)}_* \right) \setminus D^{(\geq 0)} \\
		& = & \left( \bigcup_{i=0}^L \left(F^{(i)} \cup M^{(i)}\right) \right) \setminus D^{(\geq 0)} \\
		& = & \left( \bigcup_{i=0}^L F^{(i)} \right) \setminus D^{(\geq 0)}  =  F \setminus D^{(\geq 0)}  =  F \cap E_{(a)}  =  F_{(a)}.
	\end{eqnarray*}
\end{proof}

\begin{observation}
	\label{ob:gamma:a:weight}
	$\gamma_{(a)}(e) < \beta$ for all edges $e \in F_{(a)}$. 
\end{observation}

\begin{proof}
	Consider any edge $e \in F_{(a)}$. We derive that:
	\begin{eqnarray*}
		\gamma_{(a)}(e) =  \gamma^{(L)}_*(e)  =  2^{\ell_*(e)} \cdot \lambda \leq  2^L \cdot \lambda  <  \beta.
	\end{eqnarray*}
	In the derivation above, the second inequality follows from \Cref{ob:gamma:L:level}.
\end{proof}

Finally, recall that $w_{(a)}$ is the restriction of $w$ on the active edges. Thus,  $w_{(a)}$ is a $\lambda$-uniform weight-function with support $E_{(a)}$, and we have  $w_a(e) = w(e) = \lambda$ for all $e \in E_{(a)}$.

\begin{observation}
	\label{cl:wrap:up:3}
	We have:
	\begin{enumerate}
		\item $w_{(a)}(v) \leq w(v)$ for all nodes $v \in V$.
		\item $size(w) \leq (1+\epsilon) \cdot size \left( w_{(a)} \right)$. 
	\end{enumerate}
\end{observation}

\begin{proof}
	Part-(1) of the observation holds since $w_{(a)}$ is a $\lambda$-uniform weight-function with support $E_{(a)}$, $w$ is a $\lambda$-uniform weight-function with support $E$, and $E = E_{(a)} \cup E_{(p)}$. 
	
	Next, observe that $size\left( w_{(a)} \right) = \lambda \cdot \left| E_{(a)} \right|$ and $size(w) = \lambda \cdot |E| = \lambda \cdot \left| E_{(a)} \cup E_{(p)} \right| = \lambda \cdot \left( \left| E_{(a)} \right| + \left| E_{(p)} \right| \right)$. Part-(2) of the observation now follows from Invariant~\ref{inv:insertion}.
\end{proof}

\subsubsection{Bounding the distance between $w_{(a)}$ and $\gamma^{(0)}_*$}
\label{sec:3}

\begin{claim}
	\label{cl:gamma:w}
	$size \left( w_{(a)} \right) \leq  size \left( \gamma_*^{(0)} \right)  \leq (1+\epsilon) \cdot size \left( w_{(a)} \right)$.
\end{claim}

\begin{proof}
	Since $\gamma_*^{(0)}$ is a $\lambda$-uniform weight-function with support $E^{(\geq 0)}$, we have:
	\begin{equation}
		\label{eq:cl:gamma:w:1}
		size \left( \gamma_*^{(0)} \right) = \lambda \cdot \left| E^{(\geq 0)} \right| = \lambda \cdot \left( \left| E_{(a)} \right| + \left| D^{(\geq 0)} \right| \right).
	\end{equation}
	The last equality holds since $E^{(\geq 0)} = E_{(a)} \cup D^{(\geq 0)}$. Now, from~(\ref{eq:cl:gamma:w:1}) and Invariant~\ref{inv:deletion}, we get:
	\begin{equation}
		\label{eq:cl:gamma:w:2}
		\lambda \cdot \left| E_{(a)} \right| \leq size \left( \gamma_*^{(0)} \right) \leq (1+\epsilon) \cdot \lambda \cdot  \left| E_{(a)} \right|
	\end{equation}
	Recall that $w_{(a)}$ is a $\lambda$-uniform weight-function with support $E = E_{(a)}$, and hence:
	\begin{equation}
		\label{eq:cl:gamma:w:3}
		size \left( w_{(a)} \right) = \lambda \cdot \left| E_{(a)} \right|.
	\end{equation}
	The claim follows from~(\ref{eq:cl:gamma:w:2}) and~(\ref{eq:cl:gamma:w:3}).
\end{proof}

\begin{corollary}
	\label{cor:cl:gamma:w}
	$dist_V\left(w_{(a)}, \gamma^{(0)}_* \right) \leq 2 \epsilon \cdot size \left( w_{(a)} \right)$. 
\end{corollary}

\begin{proof}
	Recall that $\gamma^{(0)}_*$ is a $\lambda$-uniform weight-function with support $E^{(\geq 0)}$, $w_{(a)}$ is a $\lambda$-uniform weight-function with support $E_{(a)}$, and $E_{(a)} \subseteq E^{(\geq 0)}$. Hence, we have $w_{(a)}(v) \leq \gamma^{(0)}_*(v)$ for all nodes $v \in V$. From  \Cref{prop:triangle:3} and  \Cref{cl:gamma:w}, we now derive that:
	$$dist_V\left( w_{(a)}, \gamma^{(0)}_* \right) = 2 \cdot \left( size \left( \gamma^{(0)}_* \right) - size \left( w_{(a)} \right)  \right) \leq 2 \epsilon \cdot size \left( w_{(a)} \right).$$
\end{proof}

\subsubsection{Bounding the distance between $\gamma^{(0)}_*$ and $\gamma^{(i)}_*$, for all $i \in [1, L]$}
\label{sec:4}

Recall that we write $x = y \pm z$ as a shorthand for $x \in [y-z, y+z]$.

\begin{claim}
	\label{cl:diff:gamma}
	For all nodes $v \in V$ and all $i \in [0, L-1]$, we have: $\gamma^{(i+1)}_*(v) = (1\pm 2 \epsilon) \cdot \gamma^{(i)}_*(v)$. 
\end{claim}

\begin{proof}
	
	Focus on round $i$, and note that the weight-function $\gamma^{(i)}_*$ is $2^i \lambda$-uniform when restricted to the set $E^{(\geq i)}_*$. Specifically, we have $\gamma^{(i)}_*(e) = 2^{i} \lambda$ for all $e \in E^{(\geq i)}_*$. During round $i$, we identify a subset of edges $\left(E^{(\geq i)} \setminus F^{(i)} \right) \subseteq E^{(\geq i)} \subseteq E^{(\geq i)}_*$, and let $E^{(\geq i+1)}_* \leftarrow \text{{\sc Degree-Split}}\left(E^{(\geq i)} \setminus F^{(i)}\right)$. Note that $E^{(\geq i+1)}_* \subseteq \left( E^{(\geq i)} \setminus F^{(i)}\right)$.  We now make a crucial observation which summarizes how the weight-function $\gamma^{(i+1)}_*$ is constructed from $\gamma^{(i)}_*$.

	\begin{observation}
		\label{ob:cl:diff:gamma}
		As we switch from the weight-function $\gamma^{(i)}_*$ to  $\gamma^{(i+1)}_*$, the following  happen:
		\begin{enumerate}
			\item The weights of the edges  $e \in E^{(\geq i+1)}_*$ get doubled. In contrast, the edges $e \in \left( E^{(\geq i)} \setminus F^{(i)} \right) \setminus E^{(\geq i+1)}_*$, which were part of the support of $\gamma^{(i)}_*$, get discarded from the support of $\gamma_*^{(i+1)}$.
			\item The weight of every other edge in the support of the weight-function remains unchanged.
		\end{enumerate}
	\end{observation}
	
	\medskip
	For the rest of the proof, consider any node $v \in V$. There are two possible cases.
	
	\medskip
	\noindent {\em Case 1: The node $v$ is {\em not} incident upon any edge $e \in E^{(\geq i)} \setminus F^{(i)}$.} In this case, Observation~\ref{ob:cl:diff:gamma} implies that for every edge $(u, v) \in \text{{\sc Support}}\left( \gamma^{(i)}_* \right)$, we have $(u, v) \in \text{{\sc Support}}\left( \gamma^{(i+1)}_* \right)$ with $\gamma^{(i+1)}_*(u, v) = \gamma^{(i)}_*(u, v)$. This implies that $\gamma^{(i+1)}_*(v) = \gamma^{(i)}_*(v)$. 
	
	\medskip
	\noindent {\em Case 2: The node $v$ is incident upon at at least one edge in $E^{(\geq i)} \setminus F^{(i)}$.} In this case, we must have $\text{deg}_{E^{(\geq i)} \setminus F^{(i)}}(v) > (1/\epsilon)$. This holds because otherwise the node $v$ will get added to the set $V^{(i)}$ and all the edges incident on $v$ will get included in the set $F^{(i)}$, and this will imply that $\text{deg}_{E^{(\geq i)} \setminus F^{(i)}}(v)$, which leads to a contradiction. To summarize, we have:
	\begin{equation}
		\label{eq:cl:diff:lemma:1}
		\text{deg}_{E^{(\geq i)} \setminus F^{(i)}}(v) > (1/\epsilon).
	\end{equation}
	Now, applying  \Cref{cl:static:deg:split:2}, we get:
	\begin{equation}
		\label{eq:cl:diff:lemma:2}
		\text{deg}_{E^{(\geq i+1)}_*}(v) =  \frac{\text{deg}_{E^{(\geq i)} \setminus F^{(i)}}(v)}{2} \pm 1.
	\end{equation}
	Next, from~(\ref{eq:cl:diff:lemma:1}) and~(\ref{eq:cl:diff:lemma:2}), we get:
	\begin{equation}
		\label{eq:cl:diff:lemma:3}
		\text{deg}_{E^{(\geq i+1)}_*}(v) = (1/2) \cdot (1 \pm 2\epsilon) \cdot \text{deg}_{E^{(\geq i)} \setminus F^{(i)}}(v).
	\end{equation}
	Finally, from~(\ref{eq:cl:diff:lemma:3}) and Observation~\ref{ob:cl:diff:gamma}, we infer that:
	$$\gamma^{(i+1)}_*(v) = (1\pm 2 \epsilon) \cdot \gamma^{(i)}_*(v).$$
	This concludes the proof of the claim.
\end{proof}

\begin{corollary}
	\label{cor:diff:gamma:1}
	For all $i \in [1, L]$, we have: $dist_V \left( \gamma^{(0)}_*, \gamma^{(i)}_* \right) \leq  4 \epsilon  i \cdot size\left( w_{(a)} \right)$. 
\end{corollary}

\begin{proof}
	Fix a node $v \in V$ and an $i \in [1, L]$.  From~(\ref{eq:bound:epsilon}),~\Cref{cl:diff:gamma} and~\Cref{cl:tool:exponential}, we get:
	\begin{eqnarray*}
		\gamma^{(i)}_*(v) =  (1\pm 2 \epsilon)^{i} \cdot \gamma^{(0)}_*(v)  = (1\pm 4 \epsilon i) \cdot \gamma^{(0)}_*(v).
	\end{eqnarray*}
	Rearranging the terms in the above inequality, we get:
	\begin{equation}
		\label{eq:cor:diff:gamma:1}
		\left| \gamma^{(i)}_*(v) - \gamma^{(0)}_*(v) \right| \leq 4 \epsilon i \cdot \gamma^{(0)}_*(v).
	\end{equation}
	Summing~(\ref{eq:cor:diff:gamma:1}) across all the nodes $v \in V$, we derive that:
	$$dist_V\left( \gamma^{(0)}_*, \gamma^{(i)}_* \right)  \leq 4\epsilon i \cdot \sum_{v \in V} \gamma^{(0)}_*(v) = 2 \epsilon i \cdot size\left(\gamma^{(0)}_*\right) \leq 4 \epsilon i \cdot size \left( w_{(a)} \right).$$
	In the above derivation, the last inequality follows  from~(\ref{eq:bound:epsilon}) and~\Cref{cl:gamma:w}.
\end{proof}

\begin{corollary}
	\label{cor:diff:gamma:2}
	For all $i \in [1, L]$, we have: $size \left( \gamma^{(i)}_* \right) \leq  \left( 1 + 3 \epsilon  i \right)  \cdot size\left( w_{(a)} \right)$.
\end{corollary}

\begin{proof}
	The corollary holds because:
	\begin{eqnarray*}
		size \left( \gamma^{(i)}_* \right)  & \leq & size\left( \gamma^{(0)}_* \right) + (1/2) \cdot dist_V \left( \gamma^{(0)}_*, \gamma^{(i)}_* \right) \qquad \ \ (\text{follows from  \Cref{prop:triangle:2}})\\
		& \leq & size\left( \gamma^{(0)}_* \right) + 2 \epsilon i \cdot size\left( w_{(a)} \right). \qquad \qquad \qquad (\text{follows from  \Cref{cor:diff:gamma:1}}) \\
		& \leq & (1+3\epsilon i) \cdot size \left( w_{(a)} \right). \qquad \qquad \qquad \qquad \qquad (\text{follows from~\Cref{cl:gamma:w}})
	\end{eqnarray*}
\end{proof}

\subsubsection{Bounding the distance between $\gamma^{(L)}_*$ and $\gamma_{(a)}$}
\label{sec:5}

\begin{claim}
	\label{cl:gamma:active}
	$dist_V \left( \gamma^{(L)}_*, \gamma_{(a)} \right) \leq 24 \epsilon L \cdot size\left( w_{(0)}\right)$. 
\end{claim}

\begin{proof}
	The weight-function $\gamma_{(a)}$ is obtained by discarding all the edges that are {\em not} active from the support of  $\gamma^{(L)}_*$. This means that $\text{{\sc Support}}\left( \gamma_{(a)} \right) \subseteq \text{{\sc Support}}\left( \gamma^{(L)}_{*} \right)$, and $\gamma_{(a)}(e) = \gamma^{(L)}_*(e)$ for all edges $e \in \text{{\sc Support}}\left( \gamma_{(a)} \right)$. Thus, $\gamma_{(a)}(v) \leq \gamma^{(L)}_*(v)$ for all nodes $v \in V$. Hence, from  \Cref{prop:triangle:3}, we infer that:
	\begin{equation}
		\label{eq:cl:gamma:active:1}
		dist_V \left(\gamma^{(L)}_*, \gamma_{(a)} \right) = 2 \cdot \left( size \left( \gamma^{(L)}_* \right) - size\left( \gamma_{(a)} \right) \right).
	\end{equation}
	We next derive that:
	\begin{eqnarray*}
		size \left( \gamma^{(L)}_* \right) - size\left( \gamma_{(a)} \right)  & = & \sum_{e \in \text{{\sc Support}}\left( \gamma^{(L)}_*\right) \cap D^{(\geq 0)}} \gamma^{(L)}_*(e) \\
		& \leq  & \sum_{i=0}^L 2^i \lambda \cdot \left| D^{(\geq i-1)} \right| \ \ \ (\text{see Observation~\ref{ob:gamma*} and \Cref{ob:pseudo:level:weight}}) \\
		& = & \lambda \cdot D^{(\geq -1)} + \sum_{i=0}^{L-1} 2^{i+1} \lambda \cdot \left| D^{(\geq i)} \right| \\
		& = & \lambda \cdot D^{(\geq 0)} + \sum_{i=0}^{L-1} 2^{i+1} \lambda \cdot \left| D^{(\geq i)} \right|  \qquad (\text{since } D^{(\geq -1)} = D^{(\geq 0)})\\
		& \leq & \sum_{i=0}^{L-1} 3 \cdot 2^i \lambda \cdot \left| D^{(\geq i)} \right| \\
		& \leq & \sum_{i=0}^{L-1} 3 \cdot 2^i \lambda \cdot \epsilon \cdot \left| E^{(\geq i)} \right| \qquad (\text{follows from Invariant~\ref{inv:deletion}}) \\
		& \leq & \sum_{i=0}^{L-1} 3 \epsilon \cdot size \left( \gamma^{(i)}_* \right) \qquad (\text{follows from Observation~\ref{ob:gamma:new:1}}) \\
		& \leq & \sum_{i=0}^{L-1} 6 \epsilon \cdot size \left( \gamma^{(0)}_* \right) \qquad (\text{follows from~(\ref{eq:bound:epsilon}) and~\Cref{cor:diff:gamma:2}}) \\
		& \leq & 6 \epsilon L \cdot size \left( \gamma^{(0)}_* \right) \\
		& \leq & 12 \epsilon L \cdot size \left( w_{(a)} \right) \qquad \qquad (\text{follows from  \Cref{cl:gamma:w}})
	\end{eqnarray*}
	\Cref{cl:gamma:active} now follows from~(\ref{eq:cl:gamma:active:1}).
\end{proof}

\subsection{Proof of  \Cref{lm:dyn:uniform:sparsify:2}}
\label{sec:lm:dyn:uniform:sparsify:2}

\begin{claim}
	\label{cl:wrap:up:1}
	$dist_V\left( w_{(a)}, \gamma_{(a)} \right) \leq 30 \epsilon L \cdot size\left( w_{(a)} \right)$.
\end{claim}

\begin{proof}
	Applying  \Cref{prop:triangle},  \Cref{cor:cl:gamma:w},  \Cref{cor:diff:gamma:1}  and  \Cref{cl:gamma:active}, we derive that:
	\begin{eqnarray*}
		dist_V\left( w_{(a)}, \gamma_{(a)} \right) & \leq & dist_V\left( w_{(a)}, \gamma^{(0)}_*  \right) + dist_V\left( \gamma^{(0)}_*, \gamma^{(L)}_* \right) + dist_V \left( \gamma^{(L)}_*, \gamma_{(a)} \right) \\
		& \leq & \left(2\epsilon + 4 \epsilon L + 24 \epsilon L \right) \cdot size \left( w_{(a)} \right) \\
		& = & 30 \epsilon L \cdot size \left( w_{(a)} \right).
	\end{eqnarray*}
\end{proof}

\begin{claim}
	\label{cl:wrap:up:2}
	There exists a weight-function $h' : F_{(a)} \rightarrow [0, 1]$ in $H_{(a)} = (V, F_{(a)})$ such that:
	\begin{enumerate}
		\item $h'(e) \leq \gamma_{(a)}(e)$ for all edges $e \in F_{(a)}$. 
		\item $h'(v) \leq w_{(a)}(v)$ for all nodes $v \in V$. 
		\item $size\left( w_{(a)} \right) \leq (1 - 45 \epsilon L)^{-1} \cdot size(h') \leq (1 + 50 \epsilon L) \cdot size(h')$. 
	\end{enumerate}
\end{claim}

\begin{proof}
	Note that $\text{{\sc Support}}\left( \gamma_{(a)} \right) = F_{(a)}$, according to Observation~\ref{ob:gamma:a:support}. Set $w' := \gamma_{(a)}$, $w'' := w_{(a)}$, $\alpha := 30 \epsilon L$, and recall  the value of $\epsilon$ as specified in~(\ref{eq:bound:epsilon}). The claim now follows from \Cref{lm:weight:function:scaling} and  \Cref{cl:wrap:up:1}.
\end{proof}

\begin{corollary}
	\label{cor:wrap:up:1}
	There is a weight-function $h' : F_{(a)} \rightarrow [0, 1]$ in $H_{(a)} = (V, F_{(a)})$ such that:
	\begin{enumerate}
		\item $h'(e) \leq \gamma_{(a)}(e)$ for all edges $e \in F_{(a)}$. 
		\item $h'(v) \leq w(v)$ for all nodes $v \in V$. 
		\item $size\left( w \right) \leq  (1+\epsilon) \cdot (1 + 50 \epsilon L) \cdot size(h') \leq (1 + 60 \epsilon L) \cdot size(h')$. 
	\end{enumerate}
\end{corollary}

\begin{proof}
	Follows from~(\ref{eq:bound:epsilon}),  \Cref{cl:wrap:up:2} and  \Cref{cl:wrap:up:3}.
\end{proof}

\Cref{lm:dyn:uniform:sparsify:2} now follows from  \Cref{cor:wrap:up:1} and Observation~\ref{ob:gamma:a:weight}.

\subsection{Proof of  \Cref{lm:dyn:uniform:sparsify:1}}
\label{sec:lm:dyn:uniform:sparsify:1}

The proof is analogous to the proof of  \Cref{lm:static:orientation}.  

We will orient the edges in $F := \bigcup_{i=0}^L F^{(i)}$ in such a way that every node gets an out-degree of at most $O(\epsilon^{-1} + \beta^{-1})$. Since $F \supseteq F_{(a)}$, this will imply \Cref{lm:dyn:uniform:sparsify:1}. Define  the level of a node $v \in V$  as: 
$\ell(v) = \max \left\{ i \in [0, L] : v \in V^{(\geq i)} \right\}$. Thus, for all $i \in [0, L]$ and $v \in V$, we have $\ell(v) = i$ iff $v \in V^{(i)}$.

\medskip

For any two  nodes $u, v \in V$ with $\ell(u) = \ell(v) = i < L$, we say that $u$ was assigned its level {\em before} $v$ iff we had $u \in V^{(i)}$ just before $v$ gets added to the set $V^{(i)}$ during   round $i$ in \Cref{sec:analogous:static:algo}.  We now define the following orientation of the graph $H = (V, F)$:
\begin{itemize}
	\item Consider any edge $(u, v) \in F$. W.l.o.g.~suppose that $\ell(u) \leq \ell(v)$. If $\ell(u) < \ell(v)$, then the edge   is orientated {\em from} $u$ {\em towards} $v$. Otherwise, if  $\ell(u) = \ell(v) = L$, then the edge  is oriented in any arbitrary direction. Finally, if $\ell(u) = \ell(v) < L$ and (say) the node $u$ was assigned its level before the node $v$, then  the edge  is oriented {\em from} $u$ {\em towards} $v$.
\end{itemize}

\medskip
\noindent Fix any node $x \in V$. Define $\text{Out}_F(x) := \{ (x, y) \in F : \text{the edge } (x, y) \text{ is oriented away from } x\}$. We will show that $|\text{Out}_F(x)| \leq O\left(\epsilon^{-1} + \beta^{-1} \right)$. This will imply \Cref{lm:dyn:uniform:sparsify:1}.

\medskip
\noindent {\bf (Case 1):} $\ell(x) = i < L$. Let $X^- \subseteq V^{(\geq i)}$ be the set of nodes in $V^{(\geq i)}$ that are assigned the level $i$ {\em before} the node $x$. In  words, the symbol $X^-$ denotes the status of the set $V^{(i)}$ just before $x$ gets added to $V^{(i)}$.  For every edge $(x, y) \in \text{Out}_F(x)$, we have $y \in V^{\geq i} \setminus X^{-}$ and $(x, y) \in E^{(\geq i)}$. Hence, it follows that $|\text{Out}_F(x)| \leq \text{deg}_{E^{(\geq i)}}(x, V^{(\geq i)} \setminus X^-) \leq \epsilon^{-1}$. 

\medskip
\noindent {\bf (Case 2):} $\ell(x)  = L$. Consider any edge $(x, y) \in \text{Out}_F(x)$. Clearly, we have: $(x, y) \in E^{(\geq L)}_* \subseteq \text{{\sc Support}}\left( \gamma^{(L)}_* \right)$, and  $\ell_*(x, y) = L$. From  \Cref{ob:gamma:L:level}, we now infer that: 
$$\gamma^{(L)}_*(x, y) = 2^{\ell_*(x, y)} \cdot \lambda = 2^L \cdot \lambda \geq \beta/2.$$
Thus,  we have: $\text{Out}_F(x) \subseteq \text{{\sc Support}}\left( \gamma^{(L)}_* \right)$ and $\gamma^{(L)}_*(x, y) \geq \beta/2$ for all $(x, y) \in \text{Out}_F(x)$.  This implies that: 
$\gamma^{(L)}_*(x) \geq   \sum_{(x, y) \in \text{Out}_F(x)} \gamma^{(L)}_*(x, y)  \geq   |\text{Out}_F(x)| \cdot (\beta/2).$
Rearranging the terms in this inequality, we get:
\begin{equation}
	\label{eq:1million}
	|\text{Out}_F(x) | \leq (2/\beta) \cdot \gamma^{(L)}_*(x).
\end{equation}
Next, we upper bound $\gamma^{(L)}_*(x)$ in terms of $\gamma^{(0)}_*(x)$ by applying \Cref{cl:diff:gamma}, which gives us:
\begin{equation}
	\label{eq:2million}
	\gamma^{(L)}_*(x) \leq (1+2\epsilon)^L \cdot \gamma^{(0)}_*(x) \leq (1+4 \epsilon L) \cdot \gamma^{(0)}_*(x) \leq (1+4 \epsilon L).
\end{equation}
In the derivation above, the second inequality follows from~(\ref{eq:bound:epsilon}) and Claim~\ref{cl:tool:exponential}, and the last inequality follows from \Cref{ob:gamma:0:weight}. From~(\ref{eq:1million}) and~(\ref{eq:2million}), we now infer that: $|\text{Out}_F(x)| \leq (2/\beta) \cdot (1+4\epsilon L) = O(\beta^{-1})$.  The last equality  again holds due to~(\ref{eq:bound:epsilon}).

\medskip
To summarize, under Case 1 we have shown that $|\text{Out}_F(x)| = O(\epsilon^{-1})$, whereas under Case 2 we have shown that $|\text{Out}_F(x)| = O(\beta^{-1})$. This concludes the proof of the lemma.

\section{Full Version of \Cref{sec:dyn:general}}
\label{app:sec:dyn:general}

Consider a dynamic setting where we get a graph $G = (V, E)$ with $|V| = n$ nodes and a ({\em not} necessarily uniform) fractional matching $w : E \rightarrow [0,1]$ in $G$ as input. An ``update'' either inserts/deletes an edge in $G$, or changes the weight $w(e)$ of an existing edge $e \in E$. We will show how to maintain a good matching-sparsifier  $S = (V, E_S)$ of $G$ with respect to $w$.

The rest of this section is organized as follows. We present our dynamic algorithm in \Cref{sec:static:general:algo}, and analyze its key properties in \Cref{sec:static:general:analysis}. The proof of \Cref{th:inf:main} is summarized in \Cref{sec:main:proof}.

\subsection{Our Dynamic Algorithm}
\label{sec:static:general:algo}

Our dynamic algorithm works in the following three steps.

\medskip
\noindent {\bf Step I: Discretizing the weight-function $w$.} Let $K$ be the largest integer $j$ such that $( \beta/n^2) \cdot (1+ \beta)^j < \beta$. From~(\ref{eq:bound:beta}), we infer that:
\begin{equation}
\label{eq:bound:K}
K = O(\log n).
\end{equation}
To discretize the interval $\left[ (\beta/n^2), \beta\right]$ in powers of $(1+\beta)$, we set  $\lambda_i := (\beta/n^2) \cdot (1+\beta)^i$ for all $i \in [0, K]$. Next, we define a new weight-function $\hat{w} : E \rightarrow [0,1]$, where for all edges $e \in E$:
\begin{equation*}
\label{eq:static:discretize}
\hat{w}(e) :=  \begin{cases} 0 & \text{ if } w(e) < \beta/n^2; \\
\lambda_i & \text{ else if }  \lambda_i \leq w(e) < \lambda_{i+1}  \text{ for some  } i \in [0, K]; \\
\lambda_K & \text{ else if } \lambda_K \leq w(e) < \beta; \\
w(e) & \text{ else if } w(e) \geq \beta.
\end{cases}
\end{equation*}
We now summarize a few simple properties of this weight-function $\hat{w}$.

\begin{corollary}
\label{cor:static:discretize:edge}
We have: $\hat{w}(e) \leq w(e) \leq (1+\beta) \cdot \hat{w}(e) + \beta/n^2$ for all edges $e \in E$ with $w(e) < \beta$, and $\hat{w}(e) = w(e)$ for all edges $e \in E$ with $w(e) \geq \beta$.  
\end{corollary}

\begin{corollary}
\label{cor:static:discretize}
$\hat{w}(v) \leq w(v) \leq (1+\beta) \cdot \hat{w}(v) + \beta/n$ for all nodes $v \in V$. 
\end{corollary}

\begin{proof}
Follows from \Cref{cor:static:discretize:edge} and the fact that the degree of any node in $G$ is  $\leq n$. 
\end{proof}

\begin{corollary}
\label{cor:static:maximal}
Consider any $0 < \beta <  1 - 2\beta$. If $w$ is a $(\beta, \beta)$-approximate maximal matching in $G = (V,E)$, then $\hat{w}$ is an $(3 \beta, \beta)$-approximate maximal matching in $G$. See \Cref{def:beta:approx} for the notion of an approximately maximal matching.
\end{corollary}

\begin{proof}
Suppose that $w$ is an $(\beta, \beta)$-approximate maximal matching in $G$. Take any edge $(u, v) \in E$, and consider  two mutually exclusive and exhaustive cases.

\medskip
\noindent {\bf Case 1:} $w(u, v) \geq \beta$. In this case, \Cref{cor:static:discretize:edge} implies that $\hat{w}(u, v) = w(u, v) \geq \beta$. 

\medskip
\noindent {\bf Case 2:} $w(u, v) < \beta$. In this case, since $w$ is a $(\beta, \beta)$-approximate maximal matching in $G$, there is some endpoint $x \in \{u, v\}$ of the edge $(u, v)$ which satisfies the following properties: $w(x, y) < \beta$ for all $(x, y) \in E$, and $w(x) \geq (1- \beta)$. Now, \Cref{cor:static:discretize:edge} implies that $\hat{w}(x, y) \leq w(x, y) < \beta$ for all edges $(x, y) \in E$, and \Cref{cor:static:discretize} implies that: 
$$\hat{w}(x) \geq \frac{w(x) - \beta/n}{(1+\beta)} \geq \frac{1- \beta - \beta/n}{1+\beta} \geq 1- 3 \beta \ \ \text{by (\ref{eq:bound:beta})}.$$  

The corollary follows from our analysis of  Case 1 and Case 2 above. 
\end{proof}

\medskip
\noindent {\bf Step II: Partitioning the input-graph $G = (V, E)$.} We next partition the edge-set $E$  of the input graph into subsets: $E_{-1}, E_0, E_1, \ldots, E_K, E_{\geq \beta}$, which are defined as follows.
\begin{eqnarray}
E_{-1}  & = &   \{ e \in E : \hat{w}(e) = 0 \}; \label{eq:static:partition:0} \\
 E_i & = &  \left\{ e \in E : \hat{w}(e) = \lambda_i \right\} \text{ for all } i \in [0, K]; \label{eq:static:partition:1} \\
 E_{\geq \beta} & = & \left\{ e \in E : \hat{w}(e) = w(e) \geq \beta \right\}. \label{eq:static:partition:2}
  \end{eqnarray}
 For every $i \in [0, K]$, define $G_i := (V, E_i)$, and  let $\hat{w}_i : E_i \rightarrow [0, 1]$ be the restriction of the fractional matching $\hat{w}$ onto the set $E_i$. Thus, $\hat{w}_i$ is a $\lambda_i$-uniform fractional matching in $G_i$, so that $\hat{w}_i(e) = \lambda_i$ for all edges $e \in E_i$. Similarly, let $\hat{w}_{\geq \beta} : E_{\geq \beta} \rightarrow [0, 1]$ be the restriction of $\hat{w}$ on the set $E_{\geq \beta}$. Hence, we have $\hat{w}_{\geq \beta}(e) = \hat{w}(e) \geq \beta$ for all edges $e \in E_{\geq \beta}$. 
 
 \medskip
 Note that in the dynamic setting we can perform Step I and Step II above {\em on the fly}.

\medskip
\noindent {\bf Step III: Sparsifying the subgraphs $G_0, \ldots, G_K$.}  Finally, for every $i \in [0, K]$, we run the algorithm {\sc Dynamic-Uniform-Sparsify}$(G_i, \lambda_i)$ from \Cref{sec:dyn:uniform} to maintain a matching-sparsifier of $G_i$ with respect to the $\lambda_i$-uniform fractional matching $\hat{w}_i$. Henceforth, we will use the subscript $i$  to distinguish any object that corresponds  to the call to {\sc Dynamic-Uniform-Sparsify}$(G_i, \lambda_i)$. For instance, the symbol $H_i = (V, F_i)$ will denote the sparsifier $H_{(a)} = (V, F_{(a)})$ maintained by {\sc Dynamic-Uniform-Sparsify}$(G_i, \lambda_i)$, so that $F_i \subseteq E_i$.

\medskip
Define $S := (V, E_S)$, where $E_S := E_{\geq \beta}\bigcup_{i=1}^K F_i$. We will show that $S$ is a good matching-sparsifier of $G$ with respect to $w$. 




\begin{claim}
\label{cl:sandwich:1}
\label{cl:sandwich:2}
For every $i \in [0, K]$, there is a fractional matching $h'_i : F_i \rightarrow [0, 1]$ in $H_i$, where:
\begin{enumerate}
\item  $h'_i(e) < \beta$ for all  $e \in F_i$.
\item  $h'_i(v) \leq \hat{w}_i(v)$ for all $v \in V$.
\item $size(h'_i) \leq size(\hat{w}_i) \leq \left(1 + 120 \epsilon \log n \right) \cdot size(h'_i)$.
\end{enumerate}
\end{claim}

\begin{proof}
The first two parts of the claim follow from part-(1) and part-(2) of \Cref{lm:dyn:uniform:sparsify:2}. 

Since $h'_i(v) \leq \hat{w}_i(v)$ for all nodes $v \in V$, we clearly have $size(h'_i) \leq size(\hat{w}_i)$. From part-(3) of \Cref{lm:dyn:uniform:sparsify:2}, we now derive that:
\begin{eqnarray*}
 size(\hat{w}_i) & \leq & \left(1 + 60 \epsilon \cdot \log \frac{\beta}{\lambda_i} \right) \cdot size(h'_i) \\
&  \leq & \left(1 + 60 \epsilon \cdot \log \frac{\beta}{(\delta/n^2)} \right) \cdot size(h'_i) \\
& \leq &(1 + 120 \cdot \epsilon \log n) \cdot size(h'_i).
\end{eqnarray*}
The last inequality follows from~(\ref{eq:bound:beta}).
\end{proof}

\begin{claim}
\label{cl:dyn:arboricity:new}
For every $i \in [0, K]$, the graph $H_i = (V, F_i)$ has arboricity at most $O(\log (n))$.
\end{claim}

\begin{proof}
Follows from \Cref{lm:dyn:uniform:sparsify:1}.
\end{proof}

\subsection{Analysis of Our Dynamic Algorithm}
\label{sec:static:general:analysis}

We  now derive three key properties of our dynamic algorithm in the lemmas below.

\begin{lemma}
\label{lm:sparsify:nonuniform}
The subgraph $S = (V, E_S)$ of $G = (V, E)$ admits  a valid fractional matching $\phi : E_S \rightarrow [0, 1]$ such that   $size(\phi) \leq size(\hat{w}) \leq \left(1+ 120 \epsilon \log n \right) \cdot size(\phi)$. 
\end{lemma}

The proof of \Cref{lm:sparsify:nonuniform} appears in \Cref{sec:lm:sparsify:nonuniform}.

\begin{lemma}
\label{lm:sparsify:nonuniform:general}
Set $\alpha := 3 \beta$. If $\hat{w}$ is an $(\alpha, \beta)$-approximately maximal matching in $G$, then: 
$$\mu(S) \geq (1/2) \cdot \mu(G) \cdot (1 - 2000 \epsilon \cdot \log n),$$ 
where $\mu(G')$ denotes the size of maximum (integral) matching in a graph $G'$. See \Cref{def:beta:approx} for the notion of an approximately maximal matching.
\end{lemma}

The proof of \Cref{lm:sparsify:nonuniform:general} appears in \Cref{sec:lm:sparsify:nonuniform:general}.

\begin{lemma}
\label{lm:sparsify:union:orientation}
The graph $S = (V, E_S)$ has arboricity at most $O\left( \log^2 n\right)$.
\end{lemma}

\begin{proof}
Consider any node $v \in V$. Since every edge $e \in E_{\geq \beta}$ has weight $\hat{w}(e) \geq \beta$, we have: 
$$\beta \cdot \text{deg}_{E_{\geq \beta}}(v) \leq \sum_{(u, v) \in E_{\geq \beta}} \hat{w}(u, v)  \leq \hat{w}(v) \leq w(v) \leq 1,$$ where the second-last inequality follows from \Cref{cor:static:discretize}. This implies that $\text{deg}_{E_{\geq \beta}}(v) \leq 1/\beta$. In words, the subgraph of $G$ induced by the edges in $E_{\geq \beta}$ has maximum degree at most $1/\beta$. 

Next, consider any index $i \in [1, K]$. According to \Cref{cl:dyn:arboricity:new}, the subgraph $H_i = (V, F_i)$ has arboricity at most $O(1/ \epsilon + 1/\beta)$. Hence, the union of all these $K$ subgraphs, taken together, has arboricity at most $O(K/ \epsilon + K/\beta)$,  where $K = O(\log n)$ according to~(\ref{eq:bound:K}). 

The lemma now follows from~(\ref{eq:bound:epsilon:new}).
\end{proof}

\subsubsection{Proof of \Cref{lm:sparsify:nonuniform}}
\label{sec:lm:sparsify:nonuniform}

Intuitively, we define $\phi : E_S \rightarrow [0, 1]$ to be the composition of  the weight-functions $h'_0, \ldots, h'_K$ and $\hat{w}_{\geq \beta}$. To be more specific, for every edge $e \in E_S$, we set:
\begin{eqnarray}
\label{eq:phi}
\phi(e) := \begin{cases}
\hat{w}(e) & \text{ if } e \in E_{\geq \beta}; \\
h'_i(e) & \text{ else if } e \in E_i \text{ for some } i \in [0, K].
\end{cases}
\end{eqnarray}	

\begin{claim}
\label{cl:phi:weight}
$\phi(v) \leq \hat{w}(v)$ for all nodes $v \in V$.
\end{claim}

\begin{proof}
Applying \Cref{cl:sandwich:1}: $\phi(v) = \hat{w}_{\geq \beta}(v) + \sum_{i=0}^K h'_i(v) \leq \hat{w}_{\geq \beta}(v) + \sum_{i=0}^K \hat{w}_i(v) = \hat{w}(v)$. 
\end{proof}

\Cref{cl:phi:weight} and \Cref{cor:static:discretize} imply that: $\phi(v) \leq \hat{w}(v) \leq w(v) \leq 1$ for all  $v \in V$. Hence, $\phi$ is a valid fractional matching in $S$.  To conclude the proof of the lemma, we observe that: 
\begin{eqnarray*}
size(\hat{w})   =    size\left( \hat{w}_{\geq \beta} \right) + \sum_{i=0}^{K} size(\hat{w}_i) \leq size\left( \hat{w}_{\geq \beta} \right) +\left(1+ 120 \epsilon \log (n) \right) \cdot  \sum_{i=0}^{K} size(h'_i) \\
 \leq  \left(1+ 120 \epsilon \log (n) \right) \cdot \left(size\left( \hat{w}_{\geq \beta} \right) + \sum_{i=0}^{K} size(h'_i) \right) = \left(1+ 120 \epsilon \log (n) \right) \cdot size(\phi).
\end{eqnarray*}
In the above derivation, the first inequality follows from part-(3) of \Cref{cl:sandwich:1}.

\subsubsection{Proof of \Cref{lm:sparsify:nonuniform:general}}
\label{sec:lm:sparsify:nonuniform:general}

We will crucially use the following structural theorem, whose proof appears in \Cref{sec:sparsifier:proof}.

\begin{theorem}
\label{th:key}
Consider any graph $G' = (V', E')$ and a fractional matching $w' : E' \rightarrow [0, 1]$ in $G'$ such that: For every edge $(u, v) \in E'$, either  $w'(u, v) < \beta$ or  $w'(u) + w'(v) \leq 1 + \beta + w'(u, v)$. Then $G'$ admits an integral matching $M' \subseteq E'$  of size $|M'|  \geq (1+\beta)^{-1} \cdot size(w')$. 
\end{theorem}

Define the fractional matching $\phi : E_S \rightarrow [0, 1]$ as in the proof of \Cref{lm:sparsify:nonuniform}.

\begin{claim}
\label{cl:phi:edge:weight}
Consider any edge $e \in E$.
\begin{enumerate}
\item If $\hat{w}(e) \geq \beta$, then \{$e \in E_S$ and $\phi(e) = \hat{w}(e) \geq \beta$\}.
\item Else if $\hat{w}(e) < \beta$, then either \{$e \notin E_S$\} or \{$e \in E_S$ and $\phi(e) < \beta$\}.
\end{enumerate} 
\end{claim}

\begin{proof}
If $\hat{w}(e) \geq \beta$, then $e \in E_{\geq \beta}$, and we have $\phi(e) = \hat{w}(e) \geq \beta$. 

For the rest of the proof, assume that $\hat{w}(e) < \beta$ and $e \in E_S$.  Then it must be the case that $e \in E_i$ for some $i \in [0, K]$, and hence  $\phi(e) = h'_i(e) < \beta$ according to \Cref{cl:sandwich:1}. 
\end{proof}

Define a new fractional matching $\phi_{< \beta} : E_S \rightarrow [0, 1]$ as follows, where for all $e \in E_S$:
\begin{equation}
\label{eq:phi:beta}
\phi_{< \beta}(e) = \begin{cases}
\phi(e) & \text{ if } \phi(e) < \beta; \\
0 & \text{ otherwise.}
\end{cases}
\end{equation}

Fix a maximum matching $M^* \subseteq E$ in the input graph $G = (V, E)$. Next, based on $M^*$, define the weight-function $\psi^* : E_S \rightarrow [0, 1]$, where for every edge $(u, v) \in E_S$, we have:
\begin{equation}
\label{eq:phi:beta}
\psi^*(u, v) = \begin{cases}
\phi_{< \beta}(u, v) & \text{ if either } (u, v) \notin M^* \text{ or } \phi_{< \beta}(u) + \phi_{< \beta}(v) \geq 1; \\
\phi_{< \beta}(u, v) +  \frac{1 - \phi_{< \beta}(u) - \phi_{< \beta}(v)}{2} & \text{ otherwise.}
\end{cases}
\end{equation}

It is easy to verify that $\psi^*$ is a valid fractional matching in $S = (V, E_S)$. This is because $\psi^*$ is obtained by taking a valid fractional matching $\phi_{<\beta}$ in $S = (V, E_S)$, and then increasing the weights on certain edges in such a manner that the sum of the weights on the endpoints of those edges do not exceed $1$.

\begin{claim}
\label{cl:key:101}
There is an integral matching $M_S \subseteq E_S$  of size $|M_S| \geq (1+\beta)^{-1} \cdot size(\psi^*)$.
\end{claim}

\begin{proof}
Note that $\psi^*$ is a valid fractional matching in $S = (V, E_S)$. Fix any edge $(u, v) \in E_S$ and consider two possible cases, as described below.

\medskip
\noindent {\em Case 1:} $\psi^*(u, v) = \phi_{< \beta}(u, v)$. In this case, since $\phi_{< \beta}(u, v) < \beta$, it follows that $\psi^*(u, v) < \beta$. 

\medskip
\noindent {\em Case 2:} $\psi^*(u, v) > \phi_{< \beta}(u, v)$. In this case, we have $\psi^*(u) + \psi^*(v) = 1$. 

\medskip
Thus, setting $G' := S$, $E' := E_S$ and $w' := \psi^*$ in \Cref{th:key}, we infer that there is an integral matching $M_S \subseteq E_S$ of size $|M_S| \geq (1+\beta)^{-1} \cdot size(\psi^*)$. 
\end{proof}

\begin{claim}
\label{cl:key:100}
$size(\psi^*) \geq (1/2) \cdot |M^*| \cdot \left(1 - (1-\alpha)^{-1} \cdot 360 \epsilon \log n \right) \cdot (1- \alpha)$. 
\end{claim}

We will shortly give a formal proof of \Cref{cl:key:100}. But now, we observe that by \Cref{cl:key:101} and \Cref{cl:key:100},  there is a matching $M_S \subseteq E_S$ in $S = (V, E_S)$ of size:

\begin{eqnarray*}
	|M_S| & \geq & (1/2) \cdot |M^*| \cdot (1+\beta)^{-1} \cdot \left(1 - (1-\alpha)^{-1} \cdot 360 \epsilon \log n  \right) \cdot (1- \alpha) \\
	& \geq & 1/2 \cdot |M^*| \cdot \frac{1 - 3\beta}{1 + \beta} \cdot \left(1 - \frac{360 \epsilon \log n}{1 - 3\beta}\right) \qquad \qquad \qquad \qquad \qquad \qquad (\text{as $\alpha = 3 \beta$})\\
	& \geq & 1/2 \cdot |M^*| \cdot  (1 - 1000 \cdot \epsilon \log n) \cdot (1 - 1000 \cdot \epsilon \log n) \qquad \qquad \qquad \text{(by (\ref{eq:bound:beta}))}\\
	& \geq & 1/2 \cdot |M^*| \cdot (1 - 2000 \cdot \epsilon \log n)\\
\end{eqnarray*}
 
Since $M^*$ is a maximum matching in $G$, this implies \Cref{lm:sparsify:nonuniform:general}. 

\medskip
\noindent {\bf Proof of \Cref{cl:key:100}:}  Say that a node $v \in V$ is {\em critical} iff: (1) $\hat{w}(v) \geq 1-\alpha$, and (2)  $\hat{w}(u, v) < \beta$ for all edges $(u, v) \in E$. Let $V_{c} \subseteq V$ denote the collection of all critical nodes in $G$. Similarly, let $M^*_c = \{ (u, v) \in M^* : \text{ either } u \in V_c \text{ or } v \in V_c\}$ denote the set of edges in $M^*$ with at least one critical endpoint. 

\begin{observation}
\label{ob:key:new:1}
$\hat{w}(u, v) \geq \beta$ for every edge $(u, v) \in E$ with $\{u, v\} \cap V_c = \emptyset$.
\end{observation}

\begin{proof}
Consider any edge $(u,v) \in E$ with $\{u, v\} \cap V_c = \emptyset$, and suppose that $\hat{w}(u, v) < \beta$. Since $\hat{w}$ is an $(\alpha, \beta)$-approximately maximal matching in $G = (V, E)$, this implies that either $u \in V_c$ or $v \in V_c$, which leads to a contradiction. 
\end{proof}

\begin{observation}
\label{ob:key:new:2}
$\hat{w}(u, v) < \beta$ for every edge $(u, v) \in E$ with $\{u, v\} \cap  V_c \neq \emptyset$.
\end{observation}

\begin{proof}
Consider any edge $(u, v) \in E$ with $\{u, v\} \cap V_c \neq \emptyset$, and suppose that $\hat{w}(u, v) \geq \beta$. Then by definition, we get $u \notin V_c$ and $v \notin V_c$, which leads to a contradiction.
\end{proof}

\begin{claim}
\label{cl:key:new:1}
$| M^* \setminus M^*_c| \leq \sum_{v \in V \setminus V_c} \psi^*(v)$.
\end{claim}

\begin{proof}
Consider any edge $(u, v) \in M^* \setminus M^*_c$. By \Cref{ob:key:new:1}, we have $\hat{w}(u,v) \geq \beta$. This implies that $(u, v) \in E_{\geq \beta} \subseteq E_S$, and hence $(u, v) \in E_S \cap M^*$. From the definition of the weight-function $\psi^*$, it now follows that $1 \leq \psi^*(u) + \psi^*(v)$. Summing this inequality over all edges in $M^* \setminus M^*_c$, we get: $|M^* \setminus M^*_c| \leq \sum_{(u, v) \in M^* \setminus M^*_c} \left(  \psi^*(u) + \psi^*(v) \right) \leq \sum_{v \in V \setminus V^*} \psi^*(v)$.
\end{proof}

\begin{claim}
\label{cl:key:new:2}
$|M^*_c| \leq (1-\alpha)^{-1} \cdot \sum_{v \in V_c} \psi^*(v) + (1-\alpha)^{-1} \cdot 360 \epsilon \log (n) \cdot |M^*|$. 
\end{claim}

\begin{proof}
Consider any node $v \in V_c$ and any edge $(u, v) \in E_S$.  Note that  $\hat{w}(u, v) < \beta$ as per  \Cref{ob:key:new:2}. As $(u, v) \in E_S$ and $\hat{w}(u,v) < \beta$, part-(2) of \Cref{cl:phi:edge:weight} implies that  $\phi(u, v) < \beta$, and hence we get: $\psi^*(u, v) \geq \phi_{< \beta} (u, v) = \phi(u, v)$. 

To summarize, for every node $v \in V_c$ and every edge $(u, v) \in E_S$, we have: $ \psi^*(u, v) \geq \phi(u,v)$. This implies that $ \psi^*(v) \geq \phi(v)$ for all nodes $v \in V_c$.  We now derive that:
\begin{eqnarray*}
\sum_{v \in V_c} \left( \hat{w}(v) -  \psi^*(v) \right) & \leq & \sum_{v \in V_c} \left( \hat{w}(v) - \phi(v) \right) \leq \sum_{v \in V} \left( \hat{w}(v) -  \phi(v) \right) \\
& = & 2 \cdot \left( size(\hat{w}) - size(\phi)\right)\leq 240 \epsilon \log (n) \cdot size(\phi).
\end{eqnarray*}
In the derivation above, the second inequality follows from \Cref{cl:phi:weight}, whereas the last inequality follows from \Cref{lm:sparsify:nonuniform}. Rearranging the terms, we now get:
\begin{equation}
\label{eq:revised:1}
\sum_{v \in V_c} \hat{w}(v) \leq \sum_{v \in V_c} \psi^*(v) + 240 \epsilon \log n \cdot size(\phi).
\end{equation}
By definition, we have $1 \leq (1-\alpha)^{-1} \cdot \hat{w}(v)$ for all nodes $v \in V_c$. This implies that:
\begin{equation}
\label{eq:revised:2}
|V_c| \leq (1-\alpha)^{-1} \cdot \sum_{v \in V_c} \hat{w}(v).
\end{equation}
Since every edge in $M^*_c$ has at least one endpoint in $V_c$, from~(\ref{eq:revised:1}) and~(\ref{eq:revised:2}) we infer that:
\begin{equation}
\label{eq:revised:3}
|M^*_c| \leq |V_c|  \leq (1-\alpha)^{-1} \cdot \sum_{v \in V_c} \psi^*(v) + (1-\alpha)^{-1} \cdot 240 \epsilon \log n \cdot size(\phi).
\end{equation}
Now, \Cref{lm:sparsify:nonuniform} guarantees that $\phi$ is a valid fractional matching in $S = (V, E_S)$. As $E_S \subseteq E$, it follows that $size(\phi) \leq \mu_f(G) \leq (3/2) \cdot \mu(G)$, where $\mu_f(G)$ (resp.~$\mu(G)$) respectively denotes  the size of the maximum fractional (resp.~integral) matching in $G = (V, E)$. Since $M^*$ is a maximum integral matching in $G$, we have $|M^*| = \mu(G)$, and hence: $size(\phi) \leq (3/2) \cdot |M^*|$. The claim follows if we combine this inequality with~(\ref{eq:revised:3}).
\end{proof}

As  $|M^*| = |M^*_c| + |M^* \setminus M^*_c|$, \Cref{cl:key:new:1} and \Cref{cl:key:new:2} imply that: $|M^*| \leq (1-\alpha)^{-1} \cdot \sum_{v \in V} \psi^*(v) + (1-\alpha)^{-1} \cdot 360 \epsilon \log (n) \cdot |M^*|$. Rearranging the terms, we get:
$$\sum_{v \in V} \psi^*(v)  \geq \left(1 - (1-\alpha)^{-1} \cdot 360 \epsilon \log n \right) \cdot (1- \alpha) \cdot |M^*|.$$
\Cref{cl:key:100} now follows from the observation that $\sum_{v \in V} \psi^*(v) = 2 \cdot size(\psi^*)$.

\subsection{Proof of~\cref{th:inf:main}}
\label{sec:main:proof}

We will show that the sparsifier $S = (V, E_S)$ maintained by our algorithm from~\Cref{sec:static:general:algo}, along with fractional matching $\phi : E_S \rightarrow [0, 1]$ in $S$ (see \Cref{lm:sparsify:nonuniform}), satisfy   \Cref{th:inf:main}.

Assume that $size(w) \geq 1$. Then summing \Cref{cor:static:discretize} over all nodes $v \in V$, we get:
$$2 \cdot size(w) \leq 2 (1+\beta) \cdot size(\hat{w}) + \beta \leq 2 (1+\beta) \cdot size(\hat{w}) + \beta \cdot size(w).$$
Rearranging the terms in the above inequality, we infer that:
$$size(w) \leq size(\hat{w}) \cdot \frac{(1 + \beta)}{(1 - \beta/2)} \leq (1 + 3 \beta) \cdot size(\hat{w}).$$
Now, from (\ref{eq:bound:beta})  and~\Cref{lm:sparsify:nonuniform}, we get:
$$size(w) \leq size(\hat{w}) \cdot (1+3\beta) \leq size(\phi) \cdot (1 + 120 \epsilon \log n) \cdot (1 + 3\beta) \leq size(\phi) \cdot (1 + \delta).$$
This proves part-(1) of \Cref{th:inf:main}.

\medskip

Next, note that if $w$ is a $(\beta,\beta)$-approximately maximal matching, then $\hat{w}$ is $(3\beta, \beta)$-approximately maximal matching, as per Corollary~\ref{cor:static:maximal}. Further, from (\ref{eq:bound:beta}), we derive that:
\begin{equation}
\label{eq:verylast:500}
\left(\frac{1}{2}\cdot (1 - 2000\epsilon \cdot \log n)\right)^{-1} \leq 2 + \delta.
\end{equation}
Hence, by applying (\ref{eq:bound:beta}), (\ref{eq:verylast:500}) and \Cref{lm:sparsify:nonuniform:general},  we get that:
$$\mu(G) \leq (2 + \delta) \cdot \mu(S).$$
This proves part-(2) of \Cref{th:inf:main}.

\medskip

Finally, part-(3) of \Cref{th:inf:main}  follows from \Cref{lm:sparsify:union:orientation}, whereas part-(4) and part-(5) of \Cref{th:inf:main} follow from \Cref{lm:dyn:uniform:update:time}.

\subsection{Proof of Theorem~\ref{th:key}}
\label{sec:sparsifier:proof}

Consider any sufficiently large positive integer $N$ (whose value will be specified later). Define a new fractional matching $\psi' : E' \rightarrow [0,1]$, where for every edge $e \in E'$:
\begin{eqnarray*}
\psi'(e) = \max \left\{\frac{i}{N} : i \text{ is a nonengative integer, and } \frac{i}{N} \leq w'(e)\right\}.
\end{eqnarray*} 
Thus, for every edge $e \in E'$, we have $\psi'(e) \leq w'(e) \leq \psi'(e) + 1/N$, and hence:
		\begin{equation}
		\label{eq:difference:discretization}
		0 \leq size(w') - size(\psi') \leq |E'|/N.
	\end{equation}
Define  $E'_{< \beta} := \{e \in E': \psi'(e) < \beta\}$,  and let $\psi'_{< \beta} : E'_{< \beta} \rightarrow [0, 1]$  be the weight-function  obtained by restricting $\psi'$ to the edges in $E'_{< \beta}$. Thus, we have $\psi'_{< \beta}(e) = \psi'(e)$ for all $e \in E'_{< \beta}$.
	
	Define a multi-graph $\mathcal{G}' = (V' ,\mathcal{E}')$ on the node-set $V'$, such that for  every edge $(u,v) \in E'$, there are $\psi'(u,v) \cdot N$ many multiedges joining the two endpoints $u$ and $v$ in $\mathcal{G}'$. Define the multigraph  $\mathcal{G}'_{< \beta} = (V', \mathcal{E}'_{< \beta})$ in a similar manner. Specifically, for every edge $(u,v) \in E'$, there are $\psi'_{< \beta}(u,v) \cdot N$ many multiedges joining the two endpoints $u$ and $v$ in $\mathcal{G}'_{< \beta}$.

For all nodes $v \in V'$, we have  $1 \geq w'(v) \geq \psi'(v) \geq \psi'_{< \beta}(v)$. This implies that $N$ is an upper bound on the maximum degree in both the multi-graphs $\mathcal{G}'$ and $\mathcal{G}'_{< \beta}$. Furthermore, since $\psi'_{< \beta}(e) < \beta$ for all $e \in E'_{< \beta}$, the edge multiplicity of $\mathcal{G}'_{< \beta}$ is at most $\beta \cdot N$. Thus, from Vizing's theorem \cite{Vizing64} we conclude that $\mathcal{G}'_{< \beta}$ admits a legal edge coloring $\chi'_{< \beta} : \mathcal{E}'_{< \beta} \rightarrow \mathcal{C}'_{< \beta}$ which uses at most $\beta \cdot N + N = (\beta+1) \cdot N$ distinct colors, that is, $\left| \mathcal{C}'_{< \beta} \right| = (\beta + 1) \cdot N$.

Next, we  will extend $\chi'_{< \beta}$ to obtain a legal edge coloring $\chi' : \mathcal{E}' \rightarrow \mathcal{C}'$ of the multigraph $\mathcal{G}'$ using two extra colors, i.e.,  $\mathcal{C}' \supseteq \mathcal{C}'_{< \beta}$ and $\left| \mathcal{C}' \right| = \left| \mathcal{C}'_{< \beta} \right| + 2 = (\beta+1)N + 2$.  Since $\chi'$ is an extension of $\chi'_{< \beta}$, we clearly have $\chi'(e) := \chi'_{< \beta}(e)$ for all multiedges $e \in \mathcal{E}'_{< \beta} \subseteq \mathcal{E}'$. In order to finish the construction of this coloring $\chi'$, it now remains to assign colors to the   multiedges in $\mathcal{E}' \setminus \mathcal{E}'_{< \beta}$. This is done by the following greedy algorithm.

The algorithm scans through the edges in $E' \setminus E'_{< \beta}$. While considering a given edge $(u, v) \in E' \setminus E'_{< \beta}$, in a greedy manner it assigns a free color to each one of the $\psi'(e) \cdot N$ many multiedges joining   $u$ and $v$. To be more specific, the algorithm works as follows.
\begin{itemize}
\item For every edge $e = (u, v) \in E' \setminus E'_{< \beta}$:
\begin{itemize}
\item Let $\mathcal{M}(e)$ denote the collection of multiedges in $\mathcal{G}'$ that correspond to the edge $e$. Each of these multiedges join the two endpoints $u$ and $v$ of $e$, and we have $|\mathcal{M}(e)| = \psi'(e) \cdot N$.
\item Let $\mathcal{E}'(v)$  (resp.~$\mathcal{E}'(u)$) denote the collection of multiedges in $\mathcal{G}'$ that are incident on $v$ (resp.~$u$). Let $\mathcal{Z}(e) := \left( \mathcal{E}'(u) \cup \mathcal{E}'(v) \right) \setminus \mathcal{M}(e)$. Since $w'(u, v) \geq \beta$, we derive that:
\begin{eqnarray}
\left| \mathcal{Z}(e) \right| & = &  \left( \psi'(u) + \psi'(v) -  \psi'(e) - \psi'(e)  \right) \cdot N  \nonumber \\
& \leq & \left( w'(u) + w'(v) -  w'(e) - \psi'(e)  \right) \cdot N + 2 \nonumber \\
& \leq & \left( 1 + \beta - \psi'(e) \right) \cdot N + 2 \nonumber \\
& = & \left( \left( \beta + 1 \right) \cdot N + 2 \right) - \psi'(e) \cdot N \nonumber \\
& = & \left| \mathcal{C}' \right| -  \left| \mathcal{M}(e) \right|. \label{eq:verylast:1}
\end{eqnarray}
\item From~(\ref{eq:verylast:1}), it follows that when we are considering the edge $e$ during this scan, we have enough free colors left in the palette $\mathcal{C}'$ to color all the  multiedges in $\mathcal{M}(e)$.
\end{itemize}
\end{itemize}
To summarize, we have derived that there is a legal edge coloring $\chi' : \mathcal{E}' \rightarrow \mathcal{C}'$ of the multigraph $\mathcal{G}'$ that uses at most $\left| \mathcal{C}' \right| = (\beta+1) N + 2$ colors. In this coloring $\chi'$, each color-class forms a valid (integral) matching in $\mathcal{G}'$, and these color-classes partition the set $\mathcal{E}'$. Thus, there exists some matching in $\mathcal{G}'$ of size at least: 
\begin{eqnarray*}
\frac{|\mathcal{E}'|}{|\mathcal{C}'|} = \frac{N \cdot size(\psi')}{|\mathcal{C}'|} \geq \frac{N \cdot size(w') - |E'|}{|\mathcal{C}'|} \geq \frac{N \cdot size(w') - |E'|}{(\beta+1) N + 2} = \frac{size(w') - \frac{|E'|}{N}}{(1+\beta) + \frac{2}{N}}. 
\end{eqnarray*}
In the derivation above, the first inequality follows from~(\ref{eq:difference:discretization}). Note that a matching in $\mathcal{G}'$ is also a matching in $G'$. Hence, we infer that there is a matching $M' \subseteq E'$ in $G'$ of size:
$$|M'| \geq \frac{size(w') - \frac{|E'|}{N}}{(1+\beta) + \frac{2}{N}}.$$
\Cref{th:key} now follows as $N$ tends to infinity.

\bibliography{paper}
\bibliographystyle{plain}

\end{document}